# Deep spectroscopy of the FUV-optical emission lines from a sample of radio galaxies at z~2.5: metallicity and ionization[*]

A. Humphrey[1,2♠], M. Villar-Martín[3], J. Vernet[4], R. Fosbury[5], S. di Serego Alighieri[6], L. Binette[1]

[1]*Instituto de Astronomía, Universidad Nacional Autónoma de México, Apartado Postal 70-264, 04510 México, DF, México*
[2]*Department of Physics, Astronomy and Mathematics, University of Hertfordshire, Hatfield, Herts, AL10 9AB, UK*
[3]*Instituto de Astrofísica de Andalucía (CSIC), Aptdo. 3004, 18080 Granada, Spain*
[4]*European Southern Observatory, Karl-Schwarzschild Str. 2, 85748 Garching-bei-München, Germany*
[5]*Space Telescope - European Coordination Facility, Karl-Schwarzschild Str. 2, 85748 Garching-bei-München, Germany*
[6]*INAF-Osservatorio Astrofisico di Arcetri, Largo Enrico Fermi 5, I-50125 Firenze, Italy*

## ABSTRACT

We present long-slit NIR spectra, obtained using the ISAAC instrument at the Very Large Telescope, which sample the rest-frame optical emission lines from nine radio galaxies at z~2.5. One-dimensional spectra have been extracted and, using broad band photometry, have been cross calibrated with spectra from the literature to produce line spectra spanning a rest wavelength of ~1200-7000 Å. The resulting line spectra have a spectral coverage that is unprecedented for radio galaxies at any redshift. We have also produced a composite of the rest-frame UV-optical line fluxes of powerful, z~2.5 radio galaxies.

We have investigated the relative strengths of Lyα, Hβ, Hα, HeII λ1640 and HeII λ4687, and we find that $A_v$ can vary significantly from object to object. In addition, we have identified new line ratios to calculate electron temperature: [NeV] λ1575/[NeV] λ3426, [NeIV] λ1602/[NeIV] λ2423, OIII] λ1663/[OIII] λ5008 and [OII] λ2471/[OII] λ3728. We calculate an average OIII temperature of 14100±800 K.

We have modelled the rich emission line spectra, and we conclude that they are best explained by AGN-photoionization with the ionization parameter U varying between objects. For shock models (with or without the precursor) to provide a satisfactory explanation for the data, an additional source of ionizing photons is required – presumably the ionizing radiation field of the AGN. Single slab photoionization models are unable to reproduce the high- and the low-ionization lines simultaneously: the higher ionization lines imply higher U than do the lower-ionization lines. This problem may be alleviated either by combining two or more single slab photoionization models with different U, or by using mixed-medium models such as those of Binette, Wilson & Storchi-Bergmann (1996). In either case, U must vary from object to object.

On the basis of NV/NIV] and NIV]/CIV we argue that, while photoionization is the dominant ionization mechanism in the EELR, shocks make a fractional contribution (~10 per cent) to its ionization. We suggest that the NV/NIV] and NIV]/CIV ratios in the broad-line region (BLR) of some quasars suggest that shock-ionization may be important in the BLR also.

We find that in the EELR of z~2 radio galaxies the N/H abundance ratio is close to its solar value. We conclude that N/H and metallicity do not vary by more than a factor of two in our sample. These results are consistent with the idea that the massive ellipticals which become the hosts to powerful AGN are assembled very early in the history of the universe, and then evolve relatively passively up to the present day.

**Key words:** galaxies: active – galaxies: high-redshift – quasars: emission lines – ultraviolet: ISM.

# 1. INTRODUCTION

Powerful radio galaxies are believed to be hosted by massive ellipticals or their progenitors (e.g. McLure et al. 1999) and, therefore, are important probes of the formation and evolution of the most massive galactic systems. They emit intense emission lines from a wide range of ionization states, across enormous spatial scales, and which show varied kinematic properties and morphologies (e.g. Fosbury et al. 1982, 1984; Tadhunter 1986; Baum et al. 1988; Tadhunter, Fosbury & Quinn 1989).

It is now widely accepted that this line emission is powered by the active galactic nucleus (AGN hereinafter), through photoionization (e.g. Fosbury et al. 1982; Villar-Martín, Tadhunter & Clark 1997) and/or as a result of shock waves driven through nebulae by the passage of the radio jets (e.g. Dopita & Sutherland 1995, 1996). However, the balance between shock-ionization and AGN photoionization has been the subject of much debate in the literature (see e.g. Tadhunter 2002). There is a general consensus that AGN photoionization must make a substantial contribution in the majority of radio galaxies: the main arguments for AGN photoionization are (i) the detection of line emitting gas beyond the radio structure in some objects, effectively ruling out shock-ionization (e.g. Tadhunter et al. 2000; Villar-Martín et al. 2002, 2003) (ii) in a number of sources the emission line gas takes the form of a pair of diametrically opposite cones, which would be a natural consequence of illumination by the collimated radiation field of the AGN (see e.g. Jackson et al. 1998) (iii) that AGN photoionization models are able to reproduce the observed emission line ratios with few free parameters (e.g. Robinson et al. 1987), (iv) that the observed emission line ratios are often inconsistent with predictions of shock models (e.g. Villar-Martín, Tadhunter & Clark 1997; Best, Röttgering & Longair 2000) and (v) that in terms of energetics, it is feasible that the extended emission line regions (EELR) are photoionized by the AGN (e.g. Fosbury et al. 1982; McCarthy et al. 1990; Villar-Martín et al. 2002, 2003).

Although AGN photoionization has been generally successful in explaining the heating of the EELR, it is becoming increasingly clear that in at least some radio galaxies, shocks contribute significantly to the heating of the extended gas. One of the most compelling arguments for this was provided by Tadhunter et al. (2000), who detected emission line filaments in Coma A and 3C 171 lying beyond the volume of any plausible ionization cone and, in the case of Coma A, circumscribing the radio lobes. Close morphological association between the radio and line emission is also observed in other objects (e.g. Solorzano-Iñarea, Tadhunter and Bland-Hawthorn 2002; van Ojik et al. 1996), often with arc-like emission line structures circumscribing the radio lobes (Clark et al. 1998; Villar-Martín et al. 1999a; Rocca-Volmerange et al. 1994; Villar-Martín et al. 1998; McCarthy, Spinrad & van Breugel 1995; Koekemoer et al. 1999).

Remarkably strong evidence for shock-ionization has also been provided by Villar-Martín et al. (1999a; see also Clark et al. 1997), who performed a kinematic decomposition of a variety of optical emission lines in PKS 2250-41 at a number of spatial positions. These authors found that the kinematically perturbed[†] gas (FWHM ≥900 kms⁻¹) associated with the radio lobes is much hotter than can be produced by photoionization models, and moreover has line ratios more consistent with shocks than with AGN photoionization. Other, rather more circumstantial, observations suggesting that shocks may contribute to the heating of the EELR are (i) relatively high O⁺⁺ temperatures (Tadhunter, Robinson & Morganti 1989; Clark et al. 1998; Solorzano-Iñarea, Tadhunter & Axon 2001) (ii) line ratios that are more consistent with shock-ionization than with AGN photoionization (Clark et al. 1997, 1998; Villar-Martín et al. 1999a; Solorzano-Iñarea, Tadhunter & Axon 2001; Inskip et al. 2002). In addition, it has been shown that shocks are an energetically viable heating mechanism for the EELR (e.g. Koekemoer 1994; Clark 1996; Villar-Martin et al. 1999a; Bicknell et al. 2000).

The emission line spectrum of the EELR depends not only on the nature of the ionizing source; it is also a strong function of the abundance of various chemical elements. In their analysis of the optical line emission from the EELR of 11 powerful, low-z radio galaxies, Robinson et al. (1987) concluded that the metallicity of the EELR is certainly greater than one tenth of solar and probably within a factor of two of solar. These authors also found that their data lie above the power-law AGN photoionization model loci in the [NII]/Hα vs. [OI]/[OIII] diagnostic diagram, leading to the supposition that nitrogen could be several times overabundant relative to solar abundances. Subsequent investigators have found further evidence to suggest non-solar nitrogen abundances (e.g. Viegas & Prieto 1992; Tadhunter, Metz & Robinson 1994; Robinson et al. 2000). There have been a number of attempts at estimating the chemical abundances in the EELR of HzRG, where the rest-frame UV lines become observable at optical wavelengths (van Ojik et al. 1994; Villar-Martín, Tadhunter & Clark 1997; De Breuck et al. 2000; Vernet et al. 2001; Iwamuro et al. 2003; Hall et al. 2004), but none managed to fully disentangle abundance effects from ionization/excitation effects. This was due mainly to the use of a relatively small number of lines (<7), several of which were resonance lines.

For nearby radio galaxies, efforts to study the EELR have been restricted almost exclusively to the rest-frame optical emission lines (e.g. Robinson et al. 1987). Although measurements of the rest-frame UV lines have been made using space-borne observatories in a few instances (e.g. Fosbury et al. 1982), only the strongest lines were detected. Conversely, for HzRG it is the rest-frame UV lines that have been studied in most detail, and studies of the rest-frame optical lines were, until recently, restricted to only the strongest lines (compare McCarthy et al. 1992; Evans 1998 with Motohara et al. 2001; Carson et al. 2001; Larkin et al. 2002; Egami et al. 2003; Iwamuro et al. 2003).

In this paper we present new high signal to noise near-infrared and optical spectroscopic observations for a sample of 11 HzRG. We combine these spectra with optical spectra previously obtained by our group (see Vernet et al. 2001), and with optical and NIR data from the literature, to produce flux calibrated spectra spanning (in most cases) Lyα through [SII] λλ6718,6733. We then use line ratios from these spectra to understand in detail (i) the metallicity of the EELR, (ii) the ionization of the EELR, and (iii) the nature of the object-to-object variations in these ratios, with the aim of better understanding the evolutionary status of the host galaxy of HzRG, and the interplay between the host and the AGN – both of which are major goals of our programme. Throughout this paper we assume a flat universe with $H_0=71$ km s⁻¹ Mpc⁻¹, $\Omega_\lambda=0.73$ and $\Omega_m=0.27$.

---

[†] We adopt the following terminology for the kinematic properties of line emitting gas: *very broad* refers to line emission from the BLR; *narrow* refers to line emission from NLR or EELR; following Villar-Martín et al. (2002) we subdivide the narrow emission into *quiescent* (FWHM ≤1000 km s⁻¹) and *perturbed* emission (FWHM ≥1000 km s⁻¹).

**Table 1.** Log of the VLT-Antu ISAAC observations. [1] source name in either the B1950 IAU position format or the 4C catalogue format, depending on which is more commonly used in the literature [2] source redshift [3] local (Galactic) reddening (Schlegel et al. 1998) [4] date of observations in dd/mm/yy format [5] J band exposure time in seconds [6] H band exposure time in seconds [7] K band exposure time in seconds [8] position angle of the slit measured in degrees North through East [9] the full width at half of the maximum of the seeing disc, in arc seconds, measured at the DIMM or measured from a field star in the slit (1138-262 and 4C-00.54) [10] the 2σ accuracy of the relative flux calibration beween the J and H band spectrum [11] the 2σ accuracy of the relative flux calibration between the H and K band spectrum [12] reference to the broad-band magnitudes used to perform the relative flux calibration between the optical and NIR spectra: a=Pentericci et al. (1999); b=Pentericci et al. (2001); c=Röttgering et al. (1995a); d=this paper; e=Knopp & Chambers (1997).

| Name | z | E(B-V) | Date | J exp. | H exp. | K exp. | PA | FWHM | J-H cal. acc. | H-K cal. acc. | UV-opt |
|------|------|--------|------|--------|--------|--------|------|------|------|------|------|
| [1] | [2] | [3] | [4] | [5] | [6] | [7] | [8] | [9] | [10] | [11] | [12] |
| 0211-122 | 2.340 | 0.021 | 27-28/11/99 | 5200 | 7200 | 7200 | 104.0 | 0.4-0.6 | 20 | 20 | a,b |
| 0529-549 | 2.575 | 0.066 | 28/11/99 | 1600 | 3600 | 3000 | 173.1 | 0.4-0.6 | 20 | 30 | – |
| 1138-262 | 2.156 | 0.040 | 27-28/4/99 | 4800 | 2400 | 2400 | 79.6 | 1.2 | 10 | 10 | – |
| 4C-00.54 | 2.360 | 0.045 | 20-21/7/00, 9/8/00 | - | 3600 | 7200 | 170.8 | 0.8 | – | 15 | a,b |
| 1558-003 | 2.572 | 0.155 | 28/4/99, 19/4/00 | - | 3600 | 3600 | 72.4 | 0.6-0.8 | 15 | 20 | c,d |
| 4C+10.48 | 2.349 | 0.102 | 16-17,21-22/7/00, 2/8/00 | 10800 | 7300 | 14400 | 69.0 | 1-1.5 | 30 | 30 | – |
| 2025-218 | 2.630 | 0.067 | 28/4/99, 10/6/00, 2/8/00 | - | 4200 | 3600 | 175.2 | 0.6-0.8 | 20 | 30 | a,b |
| 2104-242 | 2.491 | 0.057 | 13-15/6/00 | 7200 | 6600 | 7680 | 174.0 | 1-1.5 | 20 | 30 | a,b |
| 4C+23.56 | 2.479 | 0.167 | 6/9/00, 8/9/00, 15/9/00 | 10800 | 7200 | 7200 | 47.0 | 1-1.5 | 30 | 30 | e |

## 2. OBSERVATIONS AND DATA

### 2.1 ISAAC data

#### 2.1.1 Observations

The list of sources and the journal of observations are given in table 1. All sources were selected from the ultra-steep spectrum radio galaxy survey (e.g. Röttgering et al. 1995) with redshifts between ~2.1 and ~2.6 in order to observe [OII] λ3728‡, Hβ and [OIII] λλ4960,5008, and Hα in the J, H and K bands, respectively. This sample was also chosen to overlap partially with our Keck spectropolarimetric sample (Cimatti et al. 1998; Vernet et al. 2001) to obtain a continuous wavelength coverage from Lyα to Hα. Our sample is representative of radio galaxies with high radio powers, steep radio spectra ($f_\nu \sim \nu^\alpha$ where $\alpha < -1.0$), and bright UV-optical lines and continuum.

These data were obtained with the Infrared Spectrometer And Array Camera (ISAAC; Moorwood et al. 1998) at the VLT-Antu telescope during two visitor mode runs on 27-28 April 1999 (P63) and 27-28 November 1999 (P64) and one service observing mode programme between April and September 2000 (P65).

The ISAAC short-wavelength detector is a 1k × 1k Hawaii Rockwell array. The pixel scale corresponds to 0.147" pixel$^{-1}$ on the sky. We used the low resolution grating which provides a dispersion of 3.6, 4.7 and 7.1 Å pixel$^{-1}$ in the J, H and K bands respectively, and provides coverage of one complete band in a single exposure. In combination with a 1" slit, this provides a resolution $\lambda/\Delta\lambda$ of ~500 in the J and H bands and ~450 in the K band. Each observation was carried out using nodding sequences along the slit (ABBA pattern) with a nod throw of 15" and an addition small <3" random offset around each jitter box. We used detector integration times ranging from 60 s to 120 s, depending on the band. The number of integrations at each nodding position was 1 in the P65 run and 2 in the P63 and P64 runs. With the exception of 4C-00.54, the slit was oriented along the radio axis given by Röttgering et al.

‡ We use vacuum wavelengths rounded to the nearest Å.

(1995) and Carilli et al. (1997). For 4C-00.54, the slit was placed along the major axis of the UV-optical continuum emission (Pentericci et al. 1999; 2001).

#### 2.1.2 Data reduction

These data were reduced following the classic double-pass sky subtraction technique. After flat-fielding, each pair of nodding frames (consecutive A and B position) was subtracted from each other providing a first subtraction of the sky (difference A-B frame). The geometrical distortion was mapped using Ar and Xe arc calibration frames, and also using observations of a 'trace star' observed at regular spatial intervals along the slit. The difference frames were then corrected for distortion both in the spatial and the dispersion dimension; this step also included wavelength calibration and linearization. The wavelength calibration was refined using several OH sky lines to reach an accuracy of ~0.05 pixels. A negative version (B-A) of each distortion corrected frame was then shifted by the nodding distance and added to the corresponding distortion corrected A-B frame, to remove sky residuals and to produce background subtracted 2D spectrum. For each observing sequence, the 2D spectra were registered and then coadded.

The relative response curve and telluric absorption function were estimated by fitting a black-body curve to the continuum of hot telluric standard stars; these stars were observed as close as possible, both in time and in airmass, to the observing sequences.

The absolute flux calibration was performed by integrating the spectrum of several standard stars with known broad band magnitudes, and reached an accuracy of ~30 per cent. In the case of 1558-003, we refined the absolute flux calibration by scaling to match the magnitudes measured in a 3 arcsec aperture from J, H and K imaging we obtained using ISAAC during the course of the observing programme. The magnitudes for 1558-003 are J=20.96±0.10, H=19.51±0.07, and K=18.44±0.08. J, H and K images were obtained for other sources in our sample, but the signal-to-noise ratio was insufficient to perform accurate photometry.

In some cases a relatively bright object was present in the slit and allowed the *relative* calibration between the J, H and K bands to be improved to better than 10 per cent.

The extraction apertures were selected as follows. For sources for which we have access to previously published deep optical spectroscopy (e.g. Cimatti et al. 1998; Vernet et al. 2001; Villar-Martín et al. 2003), we chose the apertures to match as closely as possible those used for the optical spectra. Otherwise, we selected the extraction apertures to make the most physical sense. For instance, in the cases of 0529-549 and 1138-262 our extraction aperture covered the full observed extent of the line emission. Both 4C+10.48 and 2104-242 show two distinct knots of line emission close to the continuum centroid, and in both cases we extracted a spectrum for each knot. When coadded, the two apertures for 2104-242 correspond to the Northern extraction aperture used for the optical spectrum by Overzier et al. (2001). The one-dimensional NIR spectra are shown in Figures 1-3.

**Table 2.** Details of all the new Keck II LRISp observations. [1] source name [2] local (Galactic) reddening (Schlegel et al. 1998) [3] source redshift [4] date of observations in dd/mm/yy format [5] exposure time in seconds [6] position angle of the slit, measured North from East.

| Name [1] | z [2] | E(B-V) [3] | Date [4] | Exp. [5] | PA [6] |
|---|---|---|---|---|---|
| 0406-244 | 2.440 | 0.053 | 07/01/00 | 7200 | 134 |
| 2025-218 | 2.630 | 0.067 | 13-14/7/99 | 18240 | 175 |

### 2.1.3 Data analysis

For the purpose of fitting the emission line parameters (i.e. flux, wavelength, width) we made use of the SPECFIT program (Kriss 1994). Gaussian line profiles were used for the fits, i.e. we assume random motions inside the emission line regions are responsible for the emission profiles. When a single Gaussian function inadequately represented an emission line, an additional component was added. The continuum level was fitted using a linear function. The stability of each fit was tested by trying a variety of initial guesses. The high surface brightness regions of 2104-242 and of 4C+10.48 are resolved into two knots apparently shifted in velocity from one another. To avoid complicating unnecessarily their velocity profiles, which is especially relevant for the deblending of Hα and [NII], we extracted spectra for each of the knots, fitted them separately and then co-added the results for each source.

In order to account for the strong sky-subtraction residuals, the severity of which varies considerably across the spectrum, spectra were extracted from off-object regions of the two-dimensional frames to represent the noise properties of the on-object spectra. The resulting 'noise spectra' were used with SPECFIT to evaluate goodness of fit and to derive 1σ uncertainties for the fit parameters. Where a useful emission line was undetected, a 3σ upper limit was placed on its flux; in this case, the line flux was assumed to be less than or equal to that of a Gaussian which has a FWHM equal to that of the other lines in the same aperture, and which has a peak flux density three times higher than the RMS noise measured in a nearby, line-free spectral region. Further details of the fitting methodology are now given on a line-by-line basis.

*[NeV] λ3426, [NeIII] λ3870, Hβ and [OI] λ6302:* these lines were each fitted by a single component, with parameters that were effectively unconstrained. However, owing to their low S/N in a few objects, Hβ or [OI] make use of constraints from Hα.

*[OII] λ3727,3730:* the lines of this doublet were specified to have equal FWHM and to have a fixed wavelength separation. Their flux ratio was set arbitrarily to unity, although varying the ratio between physically reasonable values has a negligible impact on the total flux of the doublet.

*[OIII] λλ4960,5008:* this doublet was constrained to have a flux ratio of 1:3, equal FWHM and a fixed wavelength separation; additional components were incorporated when there was a clear need for such.

*Hα and [NII] λλ6550,6585:* Hα and [NII] are blended as a result of our relatively low spectral resolution of our observations (∼600 km s⁻¹) and the intrinsic velocity widths of the lines (several hundred to ∼2000 km s⁻¹). With the exception of 0211-122 for which we do not detect [NII], the [NII] λ6585 line appears as a red asymmetry in this blend and results in a combined FWHM that is about twice that of [OIII] λ5008 and other lines. We note that [OIII] λ5008 does not show a similar red-asymmetry in our sample, and thus we reject that the asymmetry in Hα+[NII] is due to kinematic structure in Hα. To deblend Hα and [NII], the [NII] doublet was specified to have a flux ratio of 1:3, fixed wavelength separation and equal FWHM; all other parameters of [NII] and those of Hα were unconstrained. An additional component to Hα was required in some instances, to represent either emission from the broad line region (BLR hereinafter; 1138-262, 1558-003, 2025-218) or emission from kinematically perturbed gas in the EELR (0211-122). Due to low S/N in [NII] in several cases, constraints from Hα were applied.

*[SII] λλ6718,6733:* the lines of this doublet were specified to have fixed separation and equal FWHM; this doublet is almost resolved but its low S/N precluded an accurate fit, and for this reason several fits were performed with the flux ratio varied between its high and low $n_e$ limits. Constraints from Hα or [NII] were applied in some cases. The fits to [OIII] and Hα+[NII] are shown in figures 4 and 5.

Since the observations were spread over several nights (except in the case of 0529-549), it is possible that the pointings and seeing conditions differed slightly between the J, H and K band observations. In the extreme case where the source did not fill the slit, inter-band velocity shifts of up to ∼500 km s⁻¹ and differences in line FWHM of up to ∼400 km s⁻¹ between the J, H and K spectra could result. For two sources (1138-262 and 4C-00.54), a bright field star within the slit allowed us to measure the seeing FWHM between the J, H and K bands. In the case of 1138-262 the source filled the slit (seeing FWHM = 1.2") in all three bands, and thus differences in velocity and FWHM between bands are either real, or are due to different pointings. During the observations of 4C-00.54, the seeing disc FWHM was 0.8", and thus velocity shifts and differences in FWHM between lines, if they are not real, could arise due to different lines having different spatial distributions. As mentioned for 1138-262, different pointings may also be responsible for the observed differences in line velocity and FWHM.

## 2.2 Keck II data

In addition to our new NIR spectra, we also make use of deep optical spectra of ten HzRG obtained using the Low Resolution Imaging Spectrometer (LRIS, Oke et al. 1995) in polarimetry mode (Goodrich et al. 1995), at the Keck II telescope. Table 2 gives the log of these observations. For seven of the sources, the spatially collapsed spectra have been presented and discussed in detail by Cimatti et al. (1998) and Vernet et al. (2001). Villar-Martín et al. (2003) also presented and analyzed the 2D spectra for eight of the sources. (See Cimatti et al. 1998 and Vernet et al. 2001 for full details of the observations and data reduction.)

Here, we present new spectra for 0406-244 and 2025-218. These data were obtained using the same observational set-up and were reduced following similar procedures (see Cimatti et al. 1998; Vernet et al. 2001 for full details). In Figure 6 we show the spatially collapsed spectra for these two sources. We also present in Figure 6 the spatially collapsed optical spectrum of 1558-003, showing the full observed (optical) wavelength range.

**Table 3.** Details of all the spectra we make use of in this paper. [1] source name [2] Telescope used to obtain the optical (rest-frame UV) spectrum [3] Instrument used to obtain the optical spectrum [4] Dimensions, in arcsec, of the extraction aperture for the optical spectra (size along slit × width of slit) [5] Position angle of the slit, measured North through East, during the optical observations [6] Reference to the optical observations [7] Telescope used to obtain the NIR spectrum [8] Instrument used to obtain the NIR spectrum [9] Dimensions, in arcsec, of the extraction aperture for the NIR spectra (size along slit × width of slit) [10] Position angle of the slit, measured North through East, during the NIR observations [11] Reference to the NIR observations. [12] Reference to borad band photometry or other data used for the cross-calibration of the optical-NIR flux scale. References: H07=this paper; B07=Broderick et al. (2007); I03=Iwamuro et al. (2003); VM03=Villar-Martín et al. (2003); E03=Egami et al. (2003); M01=Motohara et al. (2001); V01=Vernet et al. (2001); O01=Overzier et al. (2001); R97=Röttgering et al. (1997); C98=Cimatti et al. (1998); P99=Pentericci et al. (1999); P01=Pentericci et al. (2001); Mc92=McCarthy, Persson & West (1992); R95=Röttgering et al. (1995); V99b=Villar-Martín et al. (1999b); K97=Knopp & Chambers (1997). A question mark (?) indicates that a property is not known, i.e. it was not stated explicitly in the referenced paper. H. Röttgering (private communication) provided details of the aperture size and slit PA for the optical spectra of 0200+015, 0214+183, 1138-262, and 4C+10.48.

| Name [1] | z | Tel$_{opt}$ [2] | Inst$_{opt}$ [3] | Ap$_{opt}$ [4] | PA$_{opt}$ [5] | Ref$_{opt}$ [6] | Tel$_{NIR}$ [7] | Inst$_{NIR}$ [8] | Ap$_{NIR}$ [9] | PA$_{NIR}$ [10] | Ref$_{NIR}$ [11] | Cross-cal. [12] |
|---|---|---|---|---|---|---|---|---|---|---|---|---|
| 0200+015 | 2.229 | NTT | EMMI | ~6×2 | 155 | R97 | Subaru | OHS | 6×1 | 156 | I03 | - |
| 0211-122 | 2.340 | Keck II | LRISp | 4.1×1.0 | 104 | V01 | VLT | ISAAC | 4.1×1.0 | 104 | H07 | P99,P01 |
| 0214+183 | 2.130 | NTT | EMMI | ?×2 | 177 | R97 | Subaru | OHS | 3×1 | 178 | I03 | - |
| 0406-244 | 2.440 | Keck II | LRISp | 2.4×1.0 | 134 | H07 | Subaru | OHS | 3×1 | 132 | I03 | Mc92 |
| 0529-549 | 2.575 | NTT | EMMI | 2.0×1.5 | 84 | B07 | VLT | ISAAC | 2.6×1.0 | 173 | H07 | - |
| 0731+438 | 2.429 | Keck II | LRISp | 7.7×1.0 | 12 | V01 | Subaru | OHS | 7.7×1.0 | 18 | I03 | - |
| 0828+193 | 2.572 | Keck II | LRISp | 4.1×1.0 | 44 | V01 | Subaru | OHS | 3×1 | 38 | I03 | P99,I03 |
| 1138-262 | 2.156 | NTT | EMMI | ?×2 | 87 | R97 | VLT | ISAAC | 1.8×1.0 | 80 | H07 | - |
| 4C-00.54 | 2.360 | Keck II | LRISp | 4.1×1.5 | 134 | C98 | VLT | ISAAC | 4.1×1.0 | 171 | H07 | P99,P01 |
|  |  |  |  |  |  |  | Subaru | OHS | 3×1 | 133 | I03 |  |
| 1558-003 | 2.527 | Keck II | LRISp | 2.6×1.0 | 72 | VM03 | VLT | ISAAC | 2.6×1.0 | 72 | H07 | R95,H07 |
| 4C+10.48 | 2.349 | NTT | EMMI | ~16×2 | 59 | R97 | VLT | ISAAC | 5.0×1.0 | 69 | H07 | - |
| 4C+40.36 | 2.265 | Keck II | LRISp | 4.7×1.0 | 82 | V01 | VLT | ISAAC | 4.7×1 | 83 | I03 | E03,I03 |
|  |  |  |  |  |  |  | Hale; Keck I | NIRS; NIRC | ?×0.7 | 83 | E03 |  |
| 2025-218 | 2.630 | Keck II | LRISp | 1.9×1.0 | 175 | H07 | VLT | ISAAC | 2.9×1.0 | 175 | H07 | Mc92 |
| 2104-242 | 2.491 | VLT | FORS1 | 4×1 | 18 | O01 | VLT | ISAAC | 4.2×1.0 | 174 | H07 | P99,P01,V99b |
| 4C+23.56 | 2.479 | Keck II | LRISp | 4.9×1.5 | 47 | C98 | VLT | ISAAC | 5.0×1.0 | 47 | H07 | K97 |

## 2.3 Data from the literature

To enlarge our sample, we have also included in our sample several other HzRG which have high-quality NIR spectra in literature (see table 3). In selecting these sources, we used the following criteria:
(i) redshift greater than 2;
(ii) spectra in at least three of the optical, J, H or K band;
(iii) detection of Hβ;
(iv) if the optical spectrum is used, detection of CIV λ1550, HeII λ1640 and CIII] λ1908.
Rest-frame optical line spectra for sources meeting these criteria were taken from Iwamuro et al. (2003), Motohara et al. (2001) and Egami et al. (2003), and the rest-frame UV line spectra are from Röttgering et al. (1997), Overzier et al. (2001), and Broderick et al. (2007).

## 2.4 Aperture congruence

When collating a line spectrum from several different observational data sets, it is important to consider the possible effect of an incongruence between the various extraction apertures that were used. If, for example, the emission line ratios vary between different spatial positions in the EELR (e.g. Maxfield et al. 2002; Humphrey et al. 2006), then the measured line spectrum may depend on the size and position of the extraction aperture. Thus, an incongruence between apertures could be a source of errors in our collated line spectra.

With this in mind, we have tried to ensure that our ISAAC extraction apertures are congruent with our Keck apertures. In this context we define 'congruent' to mean that the apertures do not differ in size and position by more than ~1"×1". For 10 of the 15 sources in the overall sample with both optical and NIR spectra, the apertures are congruent (Table 3). In the cases of 0200+015, 0731+438 and 4C+40.36 the original Subaru NIR extraction apertures of Iwamuro et al. (2003) were significantly smaller than used for the optical spectra. To achieve congruence between the optical and UV data for these three sources, we extracted new NIR spectra from the Subaru data of Iwamuro et al. (2003; F. Iwamuro, private communication), using extraction apertures that match those used for the optical spectra (Table 3). We note that the rest-frame optical line ratios in our new extraction do not differ significantly (i.e. at the <10% level) from those presented by Iwamuro et al. (2003). Nevertheless, for consistency we will use the line measurements from our new extractions.

In the case of 0529-549, an aperture incongruence results from the optical and NIR slits having been positioned perpendicularly. We extracted from our NIR spectrum an additional spectrum using a smaller aperture (1"×1" compared to 2.6"×1"), and we find that the line ratios do not differ within the measurement errors between the two apertures. Thus, we conclude that for this source, an aperture incongruence is unlikely to have a significant impact on the analysis of the line ratios.

For two sources, namely 0214+183 and 1138-262, we do not know explicitly the size of the extraction aperture used for the optical spectrum (Röttgering et al. 1997). In order to assess the possible dependence of the rest-frame optical line ratios on the size of the extraction aperture, we have compared the ratios measured in different sized apertures: in both cases, we find that the size of the aperture does not affect the rest-frame optical line ratios by more than 10%. Therefore, we feel the uncertainty surrounding the extraction apertures for these two sources will not have any significant impact on our analysis of their line ratios.

The optical and NIR extraction apertures are incongruent for a further two sources in our sample. In the case of 4C+10.48, we have chosen to use a smaller extraction aperture(s) for our ISAAC NIR spectrum (5") than was used by Röttgering et al. (1997) for the optical spectrum (~16"). This was to increase the S/N ratio of the Hα+[NII] blend. We find that the optical line ratios involving [OII], [OIII] and the Hα+[NII] blend do not differ by more than 10% between our 5" aperture and a 16" aperture. In the case of 4C-00.54, this is because the Keck II optical spectrum was taken with the slit oriented along the radio axis (134°) whereas during the ISAAC NIR observations the slit was positioned along the major axis of the optical continuum emission (-9.2°). In this case we are unable to quantify the effect of the incongruence.

### 2.5 Optical-NIR cross calibration

To enable us to calculate ratios using lines from the optical and NIR bands (i.e. Lyα/Hβ), we have, where possible, performed a relative flux calibration between the optical and NIR spectra. First, we converted the broad band magnitudes (see Table 3) into flux densities. From the spectra we then measured the mean flux density across a wavelength range matching that of the broad band filter. Finally, the spectra were scaled such that the mean flux density across the appropriate wavelength range matches that implied by the broad band magnitude. We were able to do this for the nine sources for which we have both Keck II or VLT optical spectra and ISAAC NIR spectra. The main source of uncertainty in this method is the assumption that ratio of line emission to continuum emission is the same outside the slit as it is within the slit. If, for example, this ratio is lower outside the slit than within it, then our calibration method will result in the line flux being overestimated, while the continuum flux will be underestimated.

In all other cases, due to the fact that we did not have access to one or more of the original spectra, or to suitable photometry, we have not been able to perform an optical-NIR cross calibration. For these sources we will not consider line ratios that use both an optical and a NIR line.

## 3. RESULTS

### 3.1 General results

The extracted one-dimensional ISAAC NIR spectra are shown in Fig 1, 2 and 3. The fits to [OIII] are shown in figure 4, and in figure 5 the fits to Hα and [NII] are shown. The parameters of the rest-frame optical lines are given in table 4. We also show the fluxes of the rest-frame UV lines when we have been able to cross-calibrate them with our ISAAC spectra. The FWHM values have been corrected in quadrature for the instrumental broadening under the assumptions that (i) the instrumental profile is Gaussian and (ii) the emitting gas filled the slit; the latter assumption may not be valid in all cases, but does not profoundly affect the conclusions of this study. Note that the FWHM errors do not take into account any uncertainty in the width of the instrumental profile. The velocities are given relative to the narrow kinematic component of [OIII] $\lambda\lambda$4960,5008. The uncertainties for the line flux measurements represent the $1\sigma$ confidence interval and are a combination of the fitting uncertainties and the uncertainties in the continuum level.

The strong lines [OIII] $\lambda$4960, [OIII] $\lambda$5008 and Hα $\lambda$6565 are detected in all objects; [OII] $\lambda\lambda$3728,3729 and [NII] $\lambda$6585 are detected in most instances; the weaker lines [NeV] $\lambda$3426, [NeIII] $\lambda$3870, Hβ $\lambda$4861, [OI] $\lambda$6302 and [SII] $\lambda\lambda$6718,6733 are detected in relatively few objects. Hereinafter these lines are referred to as [OIII], Hα, [OII], [NII], [NeV], [NeIII], Hβ, [OI] and [SII], though the full notation is sometimes used in order to avoid confusion; unless otherwise stated, [OIII] and [NII] refer only to the brighter line of each doublet and [SII] refers to the sum of doublet terms. In addition, HeII refers to HeII $\lambda$1640 unless otherwise stated. Some objects show rather large velocity shifts and differences in the FWHM between lines; these differences might be real, though we cannot discard the possibility that they are due to different pointings and seeing conditions between different observations (see §2).

In three of our sample (1138-262, 1558-003 and 2025-218), we detect very broad Hα. One of these, namely 1558-003, was not previously known to be a BLRG. In the cases of 0211-122 and 2104-242, [OIII] shows a blue asymmetry in agreement with analyses of optical spectra (Humphrey et al. 2006). We detect the giant quiescent Lyα haloes of 4C+10.48 (van Ojik et al. 1997) and 2104-242 (Villar-Martín et al. 2003) in [OIII] and Hα, which shows that these haloes are ionized, rather than being neutral and acting as a Lyα mirror.

We have constructed a composite line spectrum from our cross-calibrated optical-NIR spectra. For each source we first divided the line fluxes by Hα, and then took the average of flux of each line. This is to ensure that the composite is not dominated by the brightest sources in the sample. We show the composite line spectrum in Table 5. The advantage of our composite over that of McCarthy (1993; fluxes also shown in Table 5) is that it has a larger rest-wavelength coverage and a much narrower range in z. Ratios involving only the lines at $\lambda$<2400 Å or only the lines at $\lambda$>2400 Å show good agreement between the two composites. However, in McCarthy's composite the lines at $\lambda$<2400 Å are a factor of 2-3 stronger, relative to Hβ, than they are in our composite. Best, Röttgering & Longair (2000) noted a similar discrepancy between McCarthy's composite and their z~1 radio galaxy composite. As has been discussed by Best Röttgering & Longair (2000), it seems plausible that the discrepancy is due to the large range in z covered by McCarthy's composite and the correlation between line luminosity and z.

### 3.2 Ratios between HI and HeII recombination lines

Using line ratios formed from the HI and HeII recombination lines (Table 6), we have estimated the visual extinction $A_v$ suffered by the EELR, using (in the first instance) Hα/Hβ or HeII $\lambda$1640/HeII $\lambda$4687. We have assumed that the intrinsic value for each of the above ratios is given by Case B recombination[§], and we have adopted the average interstellar extinction curve of Savage & Mathis (1979). Table 6 gives the HI and HeII ratios for our sample, along with our estimates of $A_v$.

For four sources in our sample, the Hα/Hβ and/or HeII $\lambda$1640/HeII $\lambda$4687 ratios have allowed us to place relatively sensitive constraints on $A_v$. For 1558-003 and 4C+40.36, Hα/Hβ implies $A_v$ ~0. In the case of 0828+193, for which Hα/Hβ is not available, the HeII $\lambda$1640/HeII $\lambda$4687 ratio implies $A_v \sim 0$. In each of these three sources, Ly$\alpha$/HeII $\lambda$1640, Ly$\alpha$/H$\beta$ and HeII $\lambda$1640/H$\beta$ are in good agreement with their expected Case B values, and this appears to confirm that $A_v \sim 0$ for these three sources.

---

[§] The ratios for Case B recombination, derived from a photoionization model with U=0.1, $\alpha$=-1.5, $n_H$=100 cm$^{-3}$ and solar abundances, are Hα/Hβ=2.9, Lyα/Hβ=25, HeII $\lambda$1640/HeII $\lambda$4687=7.8, HeII $\lambda$4687/Hβ=0.2, HeII $\lambda$1640/Hβ=1.6 and Lyα/HeII $\lambda$1640=7.9. Using $\alpha$=-1.0 instead of $\alpha$=-1.5 resulted in HeII $\lambda$4687/Hβ=0.4, HeII $\lambda$1640/Hβ=3.2, and Lyα/HeII $\lambda$1640=15. These ratios do not depend strongly on U.

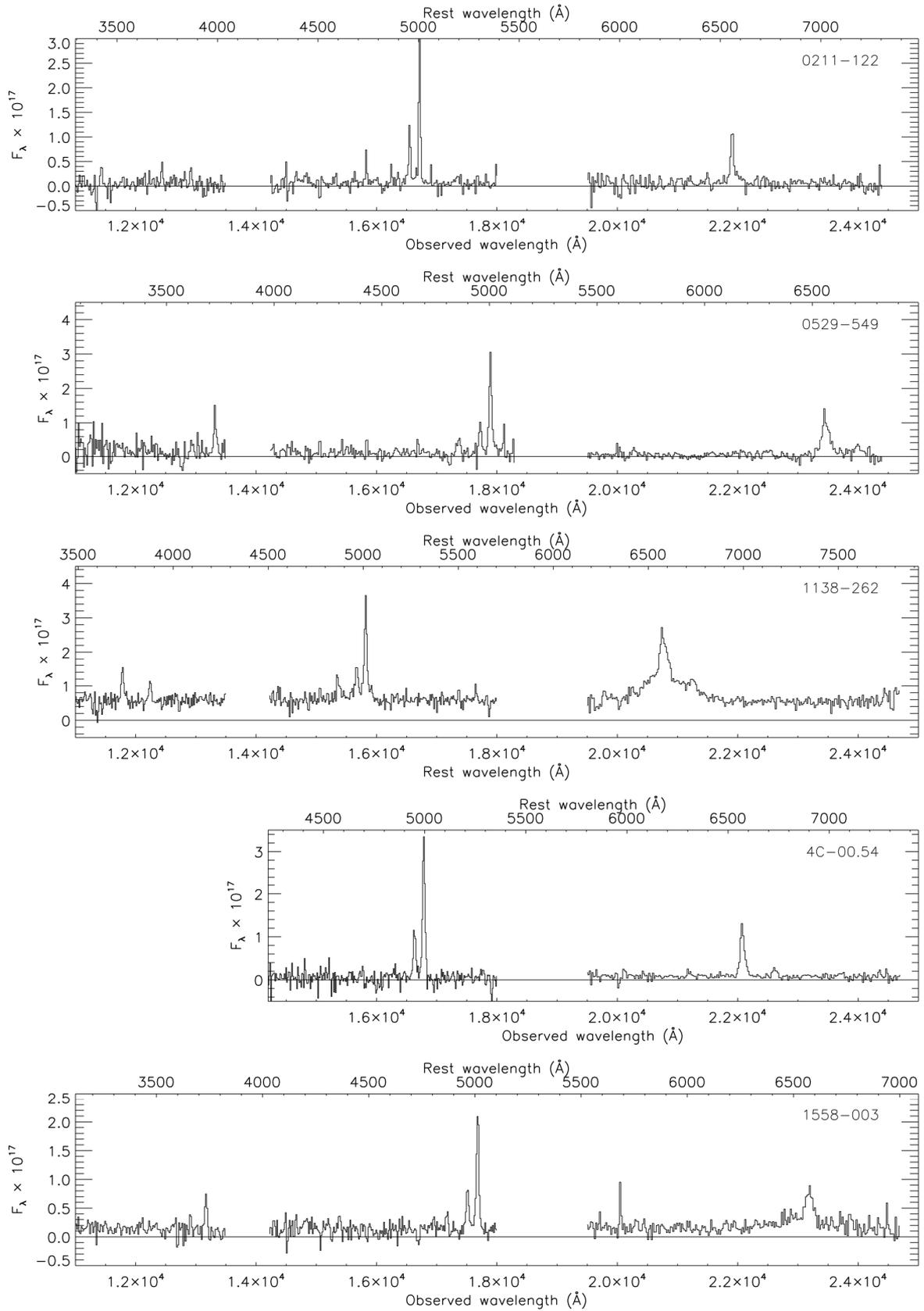

**Figure 1.** ISAAC NIR spectra of 0211-122, 0529-549, 1138-262, 4C-00.54 and 1558-003. The flux scale is in units of $10^{-17}$ erg s$^{-1}$ cm$^{-2}$ Å$^{-1}$. To reduce the prominence of sky subtraction residuals, we have binned the spectra by 3, 4 or 5 dispersion pixels and have omitted the regions near the edge of each band. In the case of 4C-00.54, we did not obtain a J-band spectrum. See text for further details of these spectra.

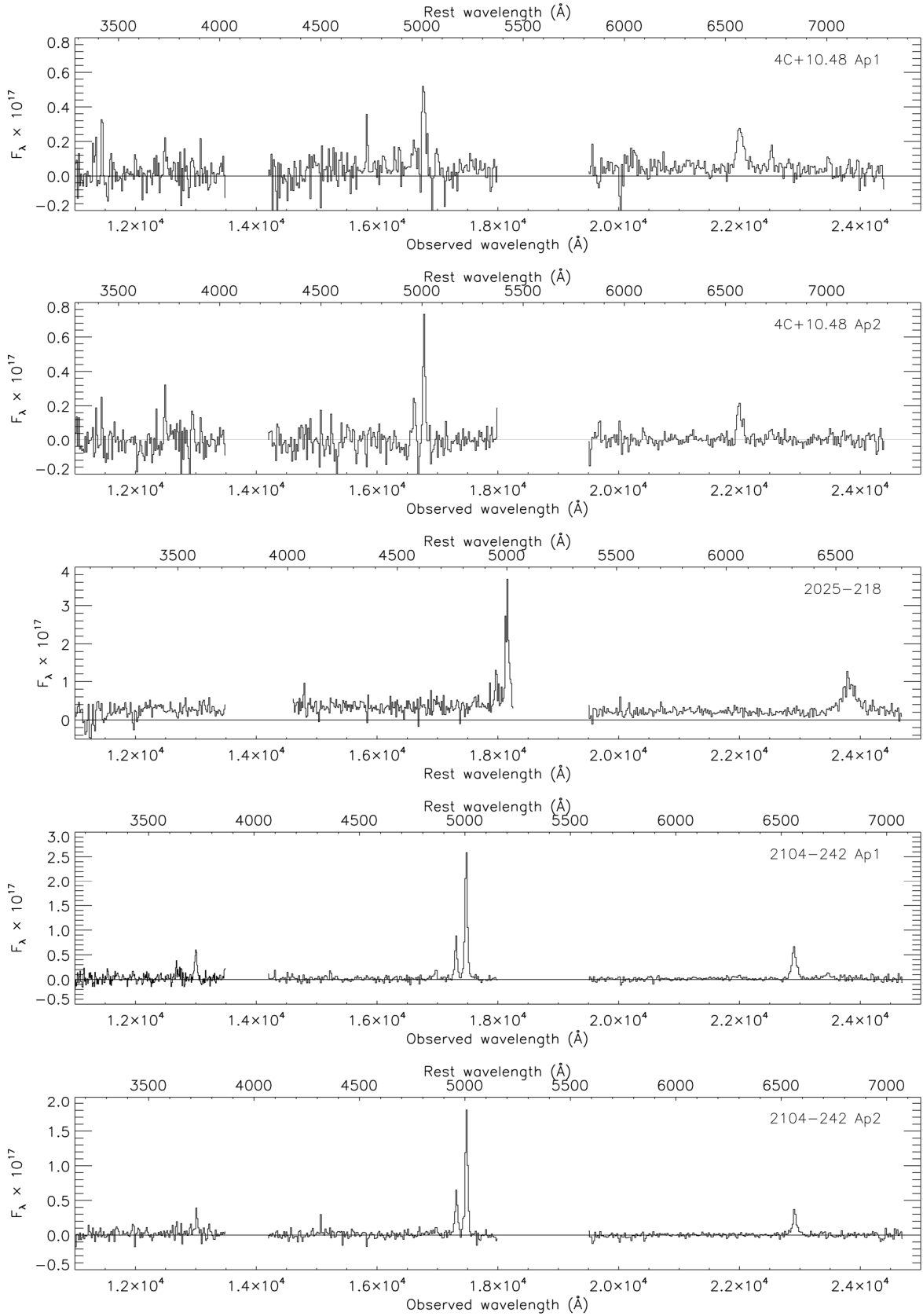

**Figure 2.** ISAAC NIR spectra of 4C+10.48, 2025-218 and 2104-242. See the caption to Fig 1 and also the text for more details.

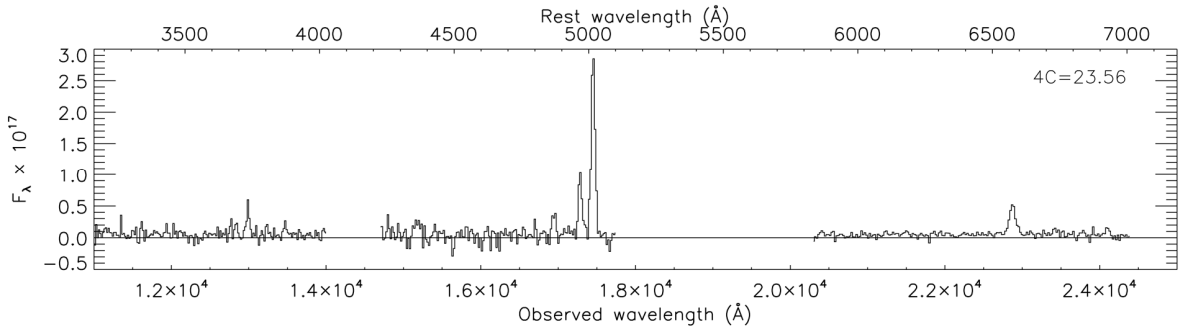

**Figure 3.** ISAAC NIR spectrum of 4C+23.56. See the caption to Fig 1 and also the text for more details.

In the case of 4C-00.54, the H$\alpha$/H$\beta$ ratio (>4.6) is significantly above its case B value and implies $A_v > 1$. The HeII $\lambda$1640/H$\beta$ ratio (0.7) is also significantly below its expected Case B value, implying $A_v$=0.3-1, and appears to confirm that the EELR of this source is significantly reddened. (4C-00.54 may be affected by the aperture incongruence: §2.4).

For 0529-549 the H$\alpha$/H$\beta$ ratio allows us to place the rather loose constraint of $A_v$ = 0.5-2.6. Unfortunately, we do not have a HeII $\lambda$1640/H$\beta$ ratio for this source.

Given that HeII $\lambda$1640/H$\beta$ implies a similar $A_v$ to that implied by H$\alpha$/H$\beta$ or HeII $\lambda$1640/HeII $\lambda$4687 (see above), we suggest that this ratio can be useful for estimating $A_v$ when HeII $\lambda$1640 or HeII $\lambda$1640/HeII $\lambda$4687 are not available. The fact that Ly$\alpha$/HeII tends to be lower in sources with lower HeII/H$\beta$ (see Table 6) is consistent with this and, furthermore, suggests that at least part of the variation in Ly$\alpha$/HeII and Ly$\alpha$/CIV between object to object (e.g. Vernet et al. 2001; Villar-Martín et al. 2007) might be due to reddening. To summarise, some of our sample appear to be moderately reddened, while others appear to be effectively unreddened. This is consistent with the results presented by Robinson et al. (1987) for low-z radio galaxies. Since $A_v$ could not be estimated accurately for the entire sample, and also because we have found evidence that $A_v$ varies from object to object, we have not applied a correction for internal extinction to the spectra. Instead, we have sought to minimise the effects of extinction on the ionization modelling through a judicious choice of line ratios.

We note that for the two sources for which we have a measurement of HeII $\lambda$4687/H$\beta$, (0406-244 and 0828+193), this ratio (0.21±0.06 and 0.29±0.05, respectively) is consistent within the errors with the value expected from photoionization models using spectral index $\alpha$ between -1.0 and -1.5 (0.4-0.2).

### 3.3 Electron temperature

The high signal to noise of the data have allowed a number of temperature sensitive line ratios to be measured, namely [NeV] $\lambda$1575/[NeV] $\lambda$3426, [NeIV] $\lambda$1602/[NeIV] $\lambda$2423, OIII] $\lambda$1663/[OIII] $\lambda$5008, [OIII] $\lambda$4364/[OIII] $\lambda$5008 and [OII] $\lambda$2471/[OII] $\lambda$3728. Their dependence on electron temperature, calculated for a fixed electron density of 100 cm$^{-3}$, is shown in figure 7. These ratios, with the exception of [NeIV] $\lambda$1602/[NeIV] $\lambda$2423, vary with $n_e$ by less than 1 per cent when densities appropriate for the EELR and narrow line region (NLR hereinafter; <20,000 cm$^{-3}$) are adopted. The [NeIV] $\lambda$1602/[NeIV] $\lambda$2423 ratio does not vary substantially (i.e. <10 per cent) if densities appropriate for the EELR (<2000 cm$^{-3}$) are used. We note that the OIII] $\lambda$1663/[OIII] $\lambda$5008 ratio is considerably more sensitive to $T_e$ than are the other ratios. The intensity ratios and resulting electron temperature estimates are given in table 7, with the uncertainties taking into account the uncertainties in the relative flux calibration and the possible effects of reddening, in addition to the uncertainties in the line flux measurements themselves. The derived temperatures are generally consistent with those measured for lower redshift radio galaxies (e.g. Tadhunter, Robinson & Morganti 1989). Using OIII] $\lambda$1663/[OIII] $\lambda$5008, we calculate an average $T_e$ of 14100±800 K. Interesting to note is the apparent temperature gradient in the EELR of 0828+193, with the temperature increasing from O$^{++}$ to Ne$^{+3}$ to Ne$^{+4}$.

### 3.4 Electron density

For one object in this sample, namely 0731+438, we have been able to place a constraint on the electron density, by comparing its [SiIII] $\lambda$1883/SiIII] $\lambda$1892 ratio with figure 2 of Keenan, Feibleman & Berrington (1992): the observed ratio of 1.7±0.2 indicates an electron density of <10$^{3.5}$ cm$^{-3}$. Thus, in this object at least, the gas responsible for the extended line emission is in the low density limit. This is consistent with density measurements in the EELR of radio galaxies at low redshift (e.g. Osterbrock 1989). Although the above ratio has also been measured for 0828+193, in this case the SiIII] $\lambda$1892 line is blended with the perturbed kinematic component of CIII] $\lambda\lambda$1907,1908 (with a FWHM of ~1700 km s$^{-1}$; see e.g. Humphrey et al. 2006), likely invalidating [SiIII]/SiIII] as an indicator of density in this object.

## 4. IONIZATION MODELLING

### 4.1 Models

To assist in understanding the origin and nature of the EELR of this sample, a number of ionization models have been either computed or sourced from the literature. Grids of single-slab photoionization models were computed using the multipurpose code MAPPINGS Ic (Binette, Dopita & Tuohy 1985; Ferruit et al. 1997), and consist of an isobaric, plane-parallel slab of gas onto which an ionizing continuum of the form $f_\nu \propto \nu^\alpha$ impinges.

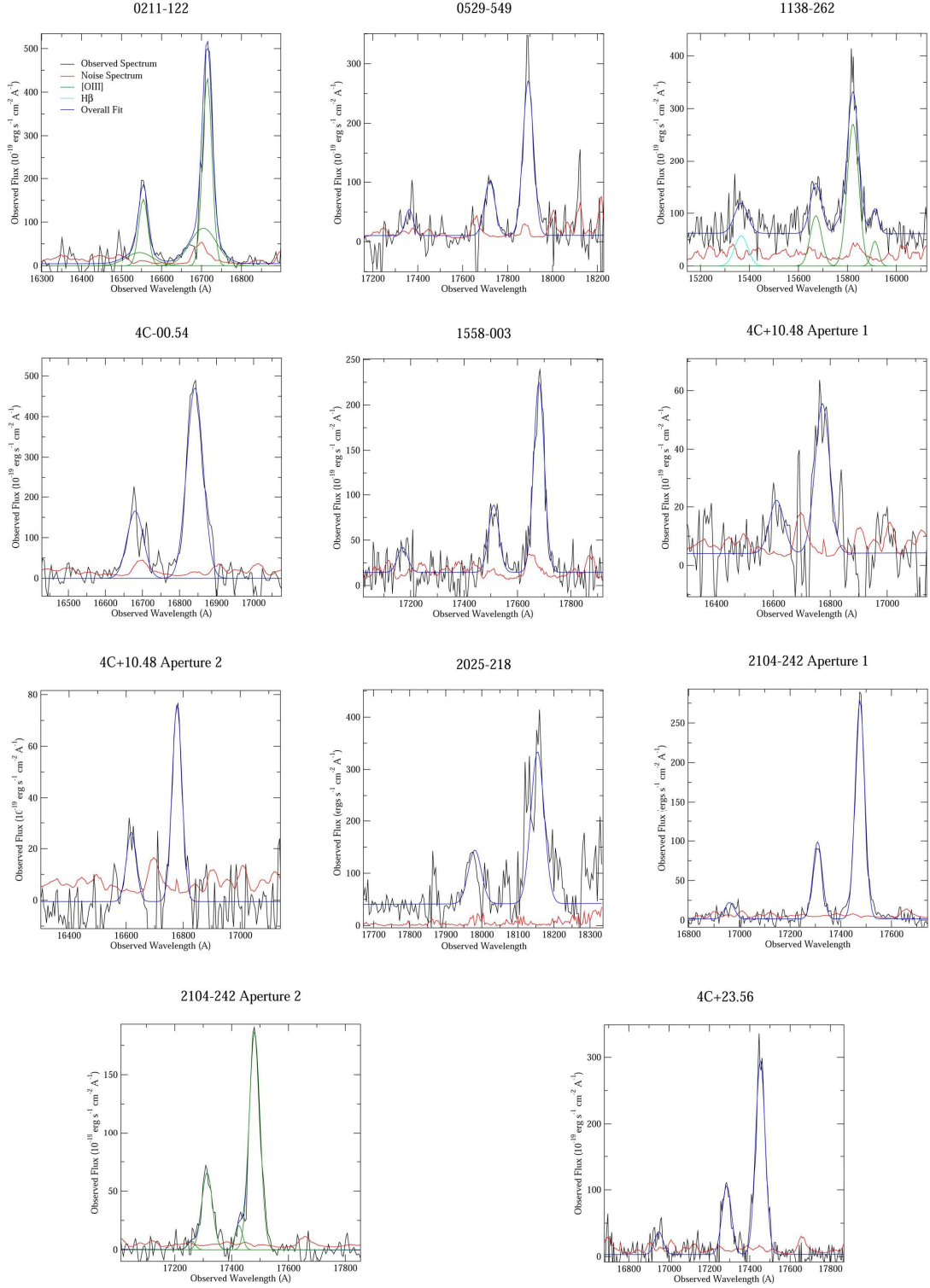

**Figure 4.** Best fits to the [OIII] λλ4960,5008 doublet and, where detected, Hβ. See the text for more details.

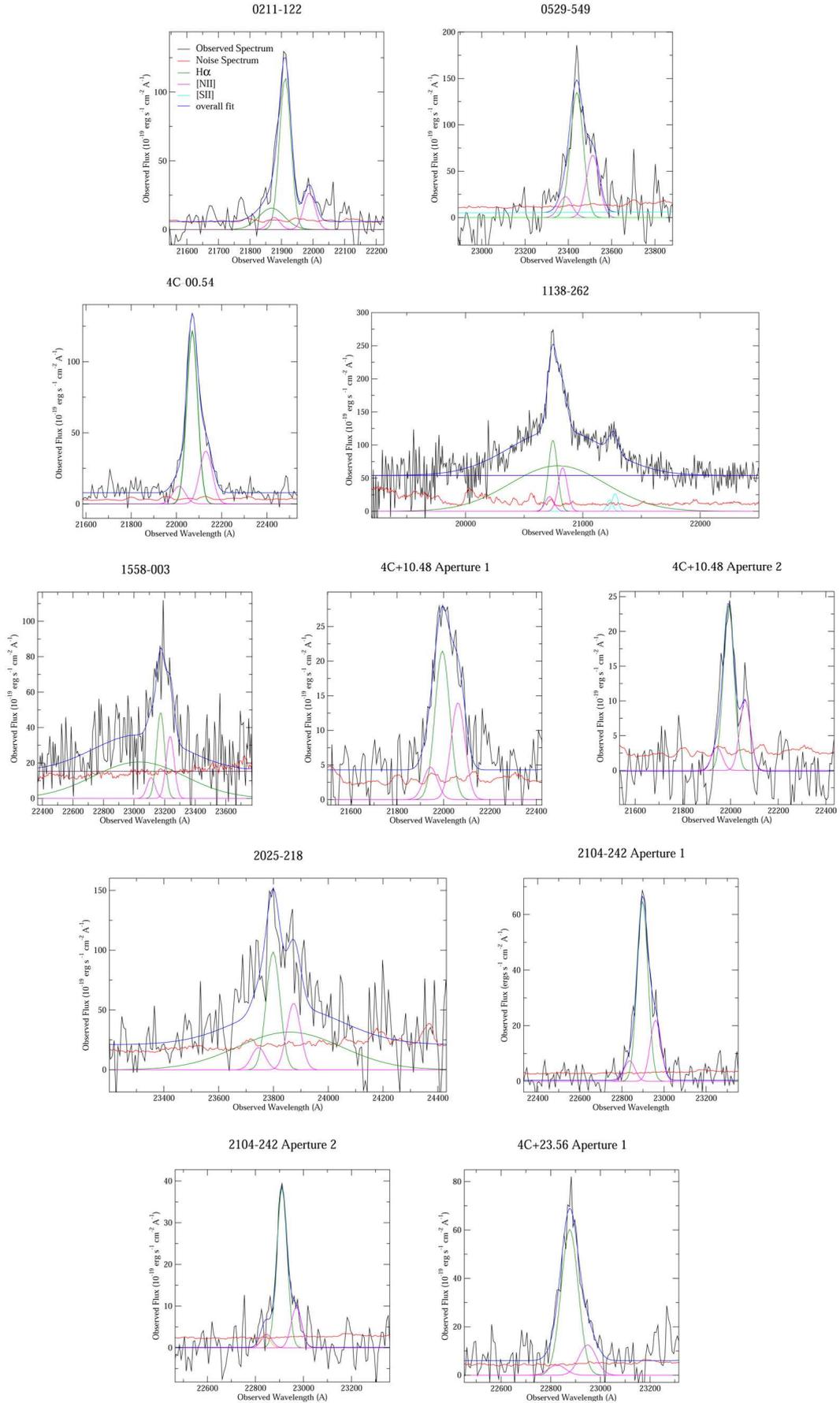

**Figure 5.** Best fits to the Hα and [NII] λλ6550,6585 doublet. The legend is shown in the first panel. See the text for more details.

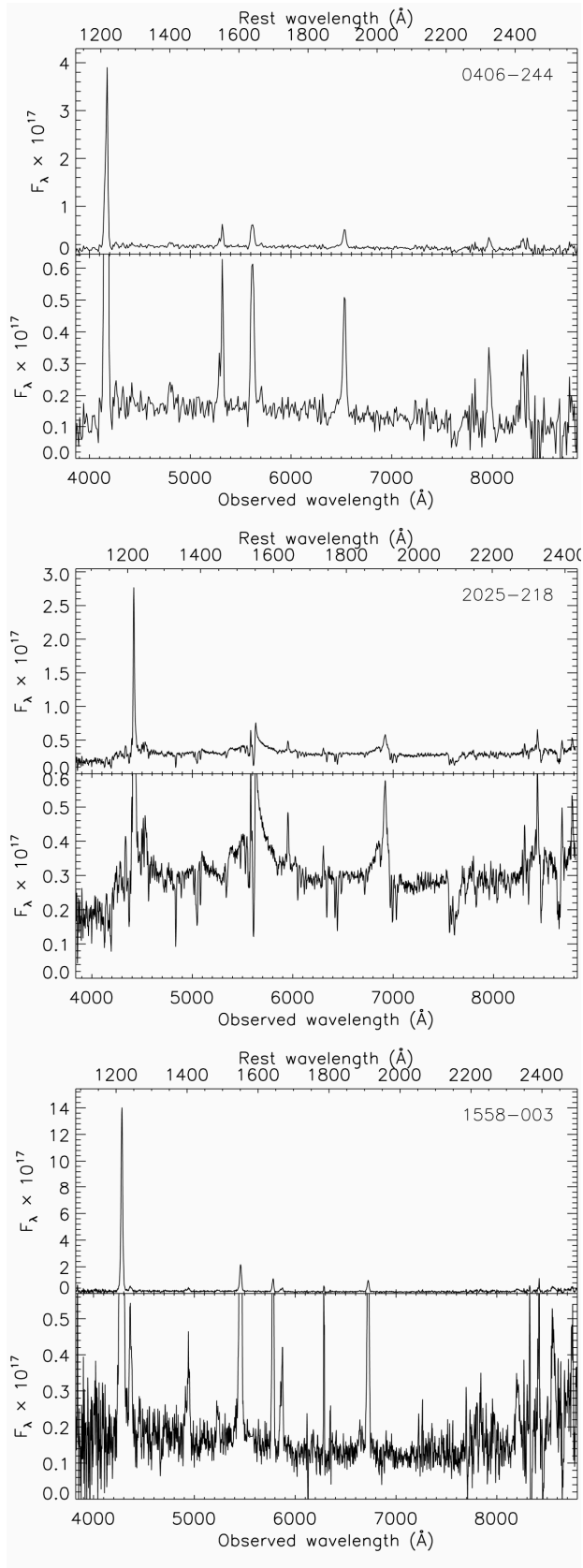

**Figure 6.** Keck II optical spectra for 0406-244, 2025-218 and 1558-003. Each spectrum is shown on two different flux scales, the first to show the strong emission lines and the second to show the continuum and weaker emission lines.

The following grids were computed:

(i) Sequence in models where ionization parameter U[**] increases from 0.0002 to 1.6 in steps of factor 2, with $\alpha$=-1.5, solar abundances (Anders & Grevesse 1989) and $n_H$=100 cm$^{-3}$. The high energy cut-off for the ionizing continuum is 50 keV to ensure that $\langle h\nu\rangle$, the mean ionizing photon energy, matches the value inferred for radio-loud AGN (e.g. Mathews & Ferland 1987).

(ii) Sequence in models as above, except that the ionizing continuum has $\alpha$=-1.0 and a 660 eV high energy truncation to ensure that $\langle h\nu$ approximately matches that of the $\alpha$=-1.5 powerlaw and the value inferred for radio-loud AGN (e.g. Mathews & Ferland 1987). We consider the two different values for $\alpha$ in order to minimize the dependence of our results on the shape of the ionizing continuum – a property that is not well known for AGN (see e.g. Haro-Corzo et al. 2007). We use both $\alpha$=-1.0 and (above) $\alpha$=-1.5 because the HeII λ4687/Hβ ratio (§3.2), where measured, implies -1.5<$\alpha$<-1.0.

(iii) Sequence in models where metallicity O/H increases from 0.25 to 4 times solar in steps of factor 2, with the abundance of each metal scaled linearly with O/H apart from nitrogen which scales quadratically with O/H (e.g. Henry, Edmunds & Köppen 2000), U=0.035, $n_H$=100 cm$^{-3}$, $\alpha$=-1.0 and a high energy cut-off of 660 eV.

(iv) Sequence in models where density increases from 10$^{-1}$ to 10$^8$ cm$^{-3}$ in steps of factor 10, with $\alpha$=-1.5, solar abundances and U=0.035.

The above models are in all cases ionization bounded, that is, the photoionization calculation is terminated when the hydrogen ionization fraction falls below 0.01. In §3.2 and §3.3, some line ratios are used to derive constraints on e.g. metallicity and density; in each instance, test calculations have been performed to ensure that such constraints are not dependent on the adoption of specific values for $\alpha$, U, metallicity or density.

One serious limitation of these models is that a range in physical conditions and different ionization mechanisms might be expected to coexist within the relatively large volume integrals contained by the spatially integrated extraction apertures (i.e. ~1000 kpc$^3$ for a 4 × 1 arc second aperture in the ±case of a z~2.5 radio galaxy) and, therefore, these models may be somewhat unrealistic. Indeed, single-slab models have encountered significant difficulties in explaining the narrow line region and extended narrow line region of active galaxies (for a detailed discussion see Robinson et al. 1987; Tadhunter, Robinson & Morganti 1989; Binette, Wilson & Storchi-Bergmann 1996). These problems have been addressed by various investigators by incorporating a range of cloud densities, the transfer of the ionizing continuum through matter-bounded followed by ionization bounded clouds, the effects of radiation pressure or simply a range in ionization parameter (e.g. Binette, Wilson & Storchi-Bergmann 1996; Binette et al. 1997; Dopita et al. 2002). We use the model of Binette, Wilson & Storchi-Bergmann (1996) to represent a mixed media photoionized nebula. This particular model considers the line emission from two distinct cloud populations: matter bounded clouds with U=0.04 and $n_H$=50 cm$^{-3}$, which only partially absorb the incident ionizing continuum, and

---

[**] Ionization parameter U is defined here as the ratio of ionizing photons to H atoms, at the face of the slab, i.e. $\frac{Q}{4\pi r^2 n_H c}$ where Q is the ionizing photon luminosity of the source, $r$ is the distance of the cloud from the ionizing source, $n_H$ is the Hydrogen density, and $c$ is the speed of light.

| Object | Hα/Hβ | $A_v$ | HeII λ1640/HeII λ4687 | $A_v$ | HeII λ1640/Hβ | $A_v$ | Lyα/Hβ | Lyα/HeII λ1640 |
|--------|-------|-------|------------------------|-------|---------------|-------|--------|-----------------|
| 0211-122 | >1.3 | unconstrained | >1.3 | <1.5 | 1.0±0.1* | 0.2-0.7 | 0.60±0.08* | 0.60±0.01 |
| 0406-244 | - | - | 6±2 | <0.5 | 1.2±0.2 | 0.2-0.4 | 7±1 | 5.8±0.7 |
| 0529-549 | 5.3±1.9 | 0.5-2.6 | - | - | - | - | - | 18±4 |
| 0828+193 | - | - | 9±2 | <0.1 | 2.6±0.4 | <0.2 | 22±3 | 8.4±0.2 |
| 1138-262 | 2.9±0.8 | <0.7 | - | - | - | - | - | 11±2 |
| 4C-00.54 | >4.2 | >1.0 | >0.7 | <2.0 | 0.7±0.1* | 0.4-1.0 | 9±1* | 12.1±0.3 |
| 1558-003 | 1.7±0.8 | 0 | >1.1 | <1.6 | 1.5±0.5 | <0.6 | 23±8 | 15.4±0.6 |
| 4C+10.48 | >2.3 | unconstrained | - | - | - | - | - | 5±1 |
| 4C+40.36 | 2.2 | 0 | 5±4 | <0.9 | 1.1±0.2 | 0.1-0.7 | 17±3 | 15±2 |
| 2104-242 | 3.3±1.1 | <1.2 | >5 | <0.4 | 3±1 | <0.2 | 43±14 | 13±2 |
| 4C+23.56 | 2.9±1.5 | <1.2 | >0.5 | <2.2 | 0.4±0.2 | 0.5-1.5 | 2.2±0.9 | 5.1±0.3 |
| Case B | 2.9 | | 7.8 | | 1.6-3.2 | | 25 | 7.9-15 |

ionization bounded clouds, which receive the partially absorbed continuum and have U=5.2 $10^{-4}$ and 2300 $cm^{-3}$. The ionizing continuum has α=-1.3 prior to impinging on the matter bounded clouds. Both cloud populations have solar abundances for the chemical elements of interest to this study. The balance between these two populations is parameterised as $A_{M/I}$, the ratio of solid angle subtended by each population at the ionizing source, and we have computed a sequence with $A_{M/I}$ = 0.0001, 0.001, 0.01, 0.04, 0.08, 0.16, 0.31, 0.63, 1.25, 2.5, 5, 10, 20, 40, 200.

An alternative to photoionization by the active nucleus is ionizing shocks, sometimes in combination with a component photoionized either by the AGN or by the shocked gas itself (e.g. Viegas & de Gouveia Dal Pino 1992; Dopita & Sutherland 1996; De Breuck et al. 2000). Here, we use the radiative shock models of Dopita & Sutherland (1996, DS96 hereinafter) to represent the possible ionizing effects of the passage of the radio structures. These models use a range of shock velocity (150, 200, 250, 300, 350, 400, 450, 500 km $s^{-1}$) and magnetic parameter (B $n^{-1/2}$ = 0, 1, 2, 4 μG $cm^{-3/2}$), and consider collisionally-ionized gas both with, and without, the photoionized precursor HII region. Limitations to these shock models are the use of a steady one-dimensional flow solution (see Allen, Dopita & Tsvetanov 1998), the neglection of the possible destruction of the post-shock, cooling clouds by the hot post-shock wind (Klein McKee & Colella 1994; Solorzano-Inarrea, Tadhunter & Axon 2001) and the use of solar abundances (whereas the less important coolants of the shocked gas depend strongly on abundances, the more important coolants do not).

## 4.2 Comparison between data and models

The one-dimensional rest-frame UV-optical spectra are now compared to the ionization models. In order to visualise this comparison, diagnostic line ratio diagrams (e.g. Baldwin, Phillips & Terlevich 1981) have been constructed. The line ratios used for the ordinates have been selected from the large number of permutations to allow the physical conditions and ionization of the EELR to be elucidated, and to minimise the impact of internal extinction on the conclusions.

Due to the fact that Hβ was not detected in most of the sample, [OIII]/Hα is used in place of the more commonly used [OIII]/Hβ. However, 0731+438 and 0828+193 do not have high quality K band spectra and, therefore, for these two sources the Hα flux has been extrapolated from Hβ assuming Hα/Hβ=2.87.

The rest-frame UV lines of 2025-218 have not been considered in this section because CIV, CIII], and possibly NV

appear to be strongly contaminated by the BLR (Villar-Martín et al. 1999b).

In the case of 0211-122 we were unable to constrain the flux of the [NII] emission from the perturbed kinematic component and, therefore, in the diagnostic diagrams we use the [NII]/Hα ratio of only the quiescent component. For all other ratios we use the sum of the two kinematic components. We feel that this will not have any significant impact on the conclusions of this work, since the [OIII]/Hα ratio does not vary significantly between the perturbed and quiescent component.

The diagnostic diagrams are shown in figure 8. The sequences in models are represented by small filled circles joined by lines, with the different sequences distinguished by the colour and pattern of the line. In some diagrams, the metallicity sequence shows a kink – this is due to the way in which $T_e$ varies with metallicity. Different sources are distinguished by symbols of different type and/or colour. For clarity, reddening vectors and error bars have been omitted, though both are considered during the analysis that follows.

To decide which models best reproduce the data, the following criteria are used. Firstly, the positions of the data in the diagnostic diagrams must be well reproduced. Second, the variation from object to object of the line ratios must also be well explained. A good level of coherence is required between the different line ratios, but considering the limitations of the ionization models it would be unreasonable to require perfect coherency. In order to avoid repetition, we hereafter refer to the rest-frame UV lines as 'the UV lines', and to the rest-frame optical lines as 'the optical lines'.

### 4.2.1 Ionization parameter sequences

Early modellers of the optical emission lines from low redshift radio galaxies found that the general trends defined by the objects, and the different regions of individual objects, can be well explained using photoionization by a hard continuum and a range in U (Robinson et al. 1987). Villar-Martín, Tadhunter & Clark (1997), the first to make serious attempt at modeling of the CIV, HeII and CIII] lines in HzRG, reached a similar conclusion, though their work was limited by the fact that only the strongest UV emission lines were available at that time.

Looking at the optical diagnostic diagrams (figure 8o-x) it is immediately apparent that photoionization by an α=-1.5 powerlaw explains satisfactorily the positions of the data. However, the picture painted by the diagrams that use only the higher-ionization UV lines is less clear (figure 8a,b,g-j). The photoionization models are able to explain most of the objects in most of these diagrams, but encounter two main problems:

**Table 7.** Electron temperatures for O⁺, O⁺⁺, Ne⁺³ and Ne⁺⁴. Error bars account for uncertainties in both the line flux measurements and the $A_v$ estimates. *In the cases of 4C-00.54 ($A_v$=1.5±0.5 mag) and 4C+23.56 ($A_v$=0.9±0.4 mag), the dereddened $\lambda\lambda$1661,1666/$\lambda$5008 ratios are used. **The rest frame optical spectrum of 1558-003 reveals it to be a broad line radio galaxy (§3.1), and thus OIII] $\lambda\lambda$1661,1666 may be contaminated by the BLR.

| Object | Species | Line ratio | Value | Implied $T_e$ (K) |
|---|---|---|---|---|
| 0211-122 | O⁺⁺ | $\lambda\lambda$1661,1666/$\lambda$5008 | $0.009^{+0.060}_{-0.002}$ | $10300^{+6300}_{-400}$ |
| 0406-244 | O⁺⁺ | $\lambda\lambda$1661,1666/$\lambda$5008 | $0.016^{+0.012}_{-0.008}$ | $11700^{+1800}_{-1500}$ |
| 0828+193 | O⁺⁺ | $\lambda\lambda$1661,1666/$\lambda$5008 | $0.043\pm0.008$ | $14800\pm700$ |
|  | O⁺⁺ | $\lambda$4364/$\lambda$5008 | $0.017\pm0.008$ | $14000\pm3000$ |
|  | Ne⁺³ | $\lambda$1602/$\lambda$2424 | $0.045\pm0.05$ | $16500\pm1000$ |
|  | Ne⁺⁴ | $\lambda$1575/$\lambda$3427 | $0.026\pm0.005$ | $22000\pm1700$ |
| 4C-00.54* | O⁺⁺ | $\lambda\lambda$1661,1666/$\lambda$5008 | $0.092^{+0.060}_{-0.043}$ | $18600^{+3600}_{-3300}$ |
| 1558-003** | O⁺⁺ | $\lambda\lambda$1661,1666/$\lambda$5008 | $0.08\pm0.02$ | $17700\pm1400$ |
| 4C+40.36 | O⁺ | $\lambda$2471/$\lambda$3728 | $0.08\pm0.02$ | $7700\pm800$ |
|  | O⁺⁺ | $\lambda\lambda$1661,1666/$\lambda$5008 | $0.017\pm0.003$ | $11800\pm500$ |
| 2104-242 | O⁺⁺ | $\lambda\lambda$1661,1666/$\lambda$5008 | $0.03\pm0.01$ | $13400\pm100$ |
| 4C+23.56* | O⁺⁺ | $\lambda\lambda$1661,1666/$\lambda$5008 | $0.039^{+0.027}_{-0.016}$ | $14400^{+2300}_{-1700}$ |

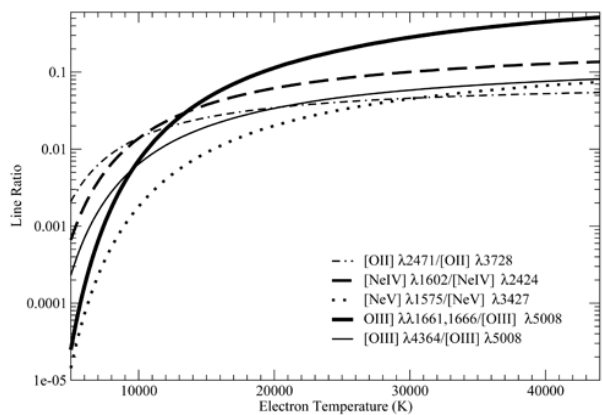

**Figure 7.** Electron temperature as a function of several emission line ratios. Notice how the OIII] $\lambda$1663/[OIII] $\lambda$5008 ratio is much more sensitive to $T_e$ than are the other line ratios in this figure.

the NIV] line is generally overpredicted relative to CIV and NV (figure 8g,i); and in the case of 0529-549, the CIII]/HeII ratio is underpredicted by a factor of ~3 (figure 8m). Both these problems were previously encountered by Vernet et al. (2001) and De Breuck et al. (2000), respectively, and will be returned to later.

Turning now to the diagrams combining low- and high-ionization lines (figure 8c,e,k,m), it can be seen that in most diagrams the data are substantially offset from the photoionization model loci. The high-ionization lines imply higher U than the lower ionization lines: for a given object the NV/CIV ratio implies higher U than CIII]/HeII, which in turn implies higher U than CII]/CIII], CII]/HeII, [OII]/[OIII] or [NII]/Hα. This result is not without precedent. McCarthy (1993) modelled a composite UV-optical line spectrum of z>0.6 radio galaxies, and found that NV and OVI are substantially underproduced by the photoionization models that well reproduce the intensities of the lower-ionization lines. He suggested that a more physical model with a range in U could provide a likely solution. The simple power-law photoionization models also fail to explain the optical high-ionization emission line ratios measured in low-z radio galaxies

(e.g. Taylor, Tadhunter & Robinson 2003). The high-ionization optical lines (e.g. HeII $\lambda$4687, [NeV], [FeVII] $\lambda$6089) are not detected from most of our high-z sources, and therefore are not included in our diagnostic diagrams.

This problem is perhaps not surprising when one considers that we are attempting to model huge volumes of ionized gas using just a single value for U. In the context of this sample, we consider an internal range in U to be the most plausible explanation.

We now consider the variation of the line ratios between objects. If there is a genuine variation in ionization level between objects, then the intensity ratio of the main coolants relative to the recombination lines, and also relative to each other, will vary significantly. Indeed, such line ratios do vary substantially in our sample: 0.2<[SII]/Hα<0.9; 0.06<[OI]/Hα<0.5; 0.2<[NeIV]/HeII<0.8; 0.2<CII]/[NeIV]<1.2; 0.1<[OIII]/[OIII]<0.4; 0.2<CII]/HeII<0.6; 0.4<[OII]/Hα<1.5; 1.5<[OIII]/Hα<4.4.

It is also important to consider whether the line ratios vary together in a manner that is consistent with a sequence in U. We find that the lower-ionization line ratios do indeed show such consistency: CII]/HeII, CIII]/HeII and [NII]/Hα increase with [OII]/[OIII]; [OIII]/Hα increases as [NII]/Hα and [OII]/[OIII] decrease (see e.g. figure 8m,n,q-x and table 8). The variation of several of the higher-ionization line ratios, such as NV/HeII, NV/CIV, CIV/HeII and CIV/CIII], with lower-ionization ratios is also consistent with a sequence in U: objects with lower [NII]/Hα have higher NV/HeII and NV/CIV (figure 8c-f and table 9); objects with lower [OII]/[OIII] have lower CIII]/HeII and higher CIV/CIII] (see e.g. figure 8k-n). As noted by Humphrey et al. (2006), sources with lower U show stronger independent evidence for jet-gas interactions.

Again, ratios involving NIV] are problematic in that they do not show any obvious relationship with other ratios, and this may suggest the presence of a second ionization mechanism in the EELR.

The measured electron temperatures are broadly consistent with those predicted by AGN photoionization models, although they do not provide strong constraints on model parameters. For instance, our model with α=-1.0, solar metallicity and U=0.1 produces OIII] $\lambda\lambda$1661,1666/$\lambda$5007 = 0.051, implying $T_e$=15500 K. For all but one of our sample, the OIII] $\lambda\lambda$1661,1666/$\lambda$5007 ratio (and hence $T_e$) is in agreement with this, within the measurement uncertainties (Table 7). In the case of 1558-003, we suggest that OIII] $\lambda\lambda$1661,1666/$\lambda$5007 is enhanced due to contamination from the BLR.

### 4.2.2 $A_{M/I}$ sequence

The $A_{M/I}$ sequence mimics the U sequence in many of the diagrams (see figure 8o,q,s,u,w). The higher-ionization lines are emitted from the matter bounded clouds, whereas the lower-ionization lines come from the ionization bounded clouds, which have relatively low U. As the balance between the two cloud populations $A_{M/I}$ varies, so does the overall ionization state. Previous investigators have found that this model sequence is able to explain the line ratios of many powerful radio galaxies (Binette, Wilson & Storchi-Bergmann 1996; Best, Röttgering & Longair 2000; De Breuck et al. 2000; Robinson et al. 2000). In particular, this model sequence has been shown to be successful in explaining simultaneously the high- and low-ionization lines from low-z radio galaxies (e.g. Taylor, Tadhunter & Robinson 2003).

We find that this model sequence explains well the positions and variation of the data in the optical diagrams (figure 8o,q,s,u,w), and also the flux of CII] relative to CIII] and CIV,

| Source | [OII]/[OIII] | CII/HeII | CIII/HeII | [NII]/Hα | [OIII]/Hα | CIV/CIII |
|---|---|---|---|---|---|---|
| 0211-122 | 0.12±0.02 | | 0.47±0.06 | <0.2 | 3.3±0.3 | 3.8±0.5 |
| 4C-00.54 | 0.15±0.03 | 0.17±0.02 | 0.45±0.03 | 0.29±0.05 | 3.4±0.3 | 3.1±0.3 |
| 2104-242 | 0.15±0.03 | | | 0.30±0.06 | 4.0±0.6 | |
| 0731+438 | 0.18±0.02 | 0.16±0.02 | 0.70±0.02 | | 4.2±0.6 | 2.2±0.07 |
| 0214-183 | 0.18±0.02 | | 1.0±0.2 | | 2.8±0.3 | 1.7±0.3 |
| 0828+193 | 0.18±0.01 | 0.19±0.02 | 0.54±0.08 | | 3.3±0.3 | 3.2±0.5 |
| 0406-244 | 0.21±0.02 | 0.50±0.09 | 0.8±0.1 | | 4.4±0.3 | 1.0±0.2 |
| 4C+23.56 | 0.26±0.06 | 0.52±0.06 | 0.67±0.08 | 0.21±0.07 | 3.4±0.3 | 1.8±0.3 |
| 1558-003 | 0.31±0.05 | 0.4±0.2 | 1.0±0.3 | 0.7±0.3 | 4.8±1.5 | 2.6±0.1 |
| 0200+015 | 0.32±0.02 | | 1.3±0.2 | | 3.8±0.7 | 1.1±0.2 |
| 1138-262 | 0.33±0.05 | | 1.0±0.3 | 0.6±0.1 | 1.5±0.2 | 0.6±0.2 |
| 4C+40.36 | 0.39±0.02 | 0.56±0.07 | 1.0±0.2 | 1.6 | 3.69±0.06 | 1.9±0.3 |
| 0529-549 | 0.4±0.1 | 1.3±0.4 | 2.7±0.7 | 0.49±0.09 | 1.5±0.2 | 1.0±0.2 |
| 4C+10.48 | 0.5±0.1 | | | 0.6±0.2 | 2.2±0.3 | |
| 53W002 | 0.66±0.07 | | 2.3±0.6 | | 2.4±0.6 | |

in agreement with Binette, Wilson & Storchi-Bergmann (1996; also e.g. Robinson et al. 2000; De Breuck et al. 2000; Taylor, Tadhunter & Robinson 2003). A further success of this model sequence is that the discrepancy between the higher-ionization lines and the lower-ionization lines (see §4.2.1) can be partially alleviated (Binette, Wilson & Storchi-Bergmann 1996). This model sequence also provides a qualitative explanation for the somewhat high CIII/HeII ratio of 0529-549.

This sequence is unable to explain the data in the diagrams involving NV and NIV] (figure 8a,c,e,g,i): in particular, the NV/HeII, NV/CIV, NV/NIV] ratios and their variation between objects are predicted to be much lower than actually observed in our sample. SiIV+OIV] and [NeIV] λ2422 vary in strength relative to NV, NIV], CIV and HeII, and this poses a similar problem. However, we feel that this need not be too much of a problem. If the matter bounded clouds have different values for U in different objects, which we feel is quite plausible considering that our sample encompasses a significant range in luminosity, then the relative strengths of NV, SiIV+OIV], NIV], CIV, HeII and [NeIV] λ2422 would then vary from object to object.

The measured electron temperatures are consistent with the range of temperatures predicted by the $A_{M/I}$ sequence.

### 4.2.3 Metallicity sequence

For a sample partially overlapping with this one, a sequence in metallicity was shown by Vernet et al. (2001) to reproduce the correlation between NV/HeII and NV/CIV, the correlation between NIV]/HeII and NIV]/CIV, and the positions of the data in many of the ultraviolet diagnostic diagrams. De Breuck et al. (2000) studied a larger sample of HzRG and also argued for a sequence in metallicity, based on the variation of NV/CIV and NV/HeII. However, the emission line ratios examined by

| Source | [NII]/Hα | NV/HeII | NV/CIV |
|---|---|---|---|
| 0211-122 | <0.2 | 0.97±0.09 | 0.54±0.05 |
| 4C+23.56 | 0.21±0.07 | 0.73±0.08 | 0.61±0.09 |
| 4C-00.54 | 0.29±0.05 | 0.59±0.04 | 0.43±0.04 |
| 2104-242 | 0.30±0.06 | 0.32±0.04 | 0.4±0.06 |
| 1138-262 | 0.6±0.1 | 0.23±0.04 | 0.38±0.09 |
| 1558-003 | 0.7±0.3 | 0.5±0.2 | 0.17±0.04 |
| 4C+40.36 | 1.6 | 0.38±0.08 | 0.22±0.04 |

Vernet et al. and De Breuck et al. did not allow the degeneracy between metallicity and ionization to be broken.

An examination of the diagnostic diagrams presented here reveals that while the sequence in metallicity models can explain positions and variation of the data in a few individual diagrams (figure 8a,i), it fails to reproduce the data in many other diagrams. Particular problems are (i) the large scatter of the data perpendicular to the metallicity sequence (all of figure 8 except panels a and i), (ii) the fact that in this sample [NII]/Hα decreases as NV/CIV and NV/HeII increase (figure 8c-f), (iii) the large variation in this sample of NV/NIV] (figure 8g), [OII]/[OIII] (figures 8k-n,q,r,w,x), [OI]/Hα (figure 8o) and [SII]/Hα (figure 8s), and (iv) the observed correlation between [OII]/[OIII] and CIII]/HeII (figure 8m).

In figure 8c,e,m,q,s,u the metallicity sequence runs roughly perpendicularly to the U sequence. If we assume that AGN photoionization is the dominant ionization mechanism (as we find later), these figures can be used to constrain the possible range in metallicity, independently of U. From the scatter parallel to the metallicity sequence, we conclude that metallicity varies by a factor of <2, as also found at low redshift by Robinson et al. (1987). As noted above, in figure 8q,s,u the data lie along the ionization parameter sequences, which have solar metallicity. We also find that for each of the five sources for which we have measurements for NV and [NeV] (i.e. 0731+438, 0828+193, 1558-003, 4C+40.36 and 4C+23.56), the NV/HeII, NV/CIV, [NeV]/[NeIII] and [NeV]/[OIII] ratios are well explained using solar metallicity and a single U value. Therefore, we conclude that the EELR of our sample have around solar N/H.

As a consistency check, we can also use the NV/HeII ratio to place a lower limit on the N/He (or N/H) abundance ratio, provided one assumes the emitting gas is photoionized by an AGN-like hard continuum (see Villar-Martín et al. 1999b). In figure 8a it can be seen that the NV/HeII ratio measured for 0211-122 corresponds to the maximum value that can be reached by the solar metallicity photoionization models. Therefore, if the EELR of 0211-122 is photoionized, then at least solar N/He is required to explain the observed NV/HeII ratio. Similarly, several other sources require at least half solar N/He. The positions of the data in figure 8c,e also suggest at least solar N abundance. The results of this consistency check adds further weight to our conclusion (above) that the EELR have roughly solar metallicity and N abundance.

If AGN photoionization is assumed to be the dominant ionization mechanism then the NIII] λ1749 line, detected from three of our sample (0828+193, 0943-242 and 4C+40.36; see Vernet et al. 2001), can also be used to constrain the N abundance. We find that NIII]/CIII], NIII]/OIII] and NIII]/HeII require around solar or super-solar N/C, N/O and N/He, respectively. The fact that these three objects have relatively low values for NV/CIV and NV/HeII, while their NIII] ratios appear to require at least solar N abundance, reinforces our conclusion that the low NV ratios are the result of low ionization state, and not of low N abundance.

Interesting to note is that when plotted against many other ratios, the [NII]/Hα ratio of 4C+40.36 is anomalously high (figure 8c,e,q,u). We conclude that this is the result of a low ionization state rather than a high N/H ratio because [SII]/Hα, which is rather insensitive to metallicity, is correspondingly high (see figure 8s).

The measurements of the electron temperatures are not inconsistent with a sequence in metallicity but do not provide any useful constraints.

### 4.2.4 Density sequence

Although this model sequence is able to reproduce consistently the observed variation in [SII]/Hα, [NII]/Hα, and [OII]/[OIII], on balance it does not explain well the data. The main problems are that contrary to the model sequence prediction (i) CIII]/HeII is observed to be correlated with [OII]/[OIII], (ii) CII]/[NeIV] is observed to increase with [OII]/[OIII], (iii) [OI]/[OIII] is observed to be lower in objects with lower [OII]/[OIII] and [NII]/Hα, (iv) [NII]/Hα is observed to decrease as NV/HeII and NV/CIV (and NV/CIV+NV/HeII) increase, and (v) NV/CIV and NIV]/CIV are observed to vary dramatically. The electron temperature measurements are not inconsistent with a density sequence.

### 4.2.5 Pure shocks

An argument against shock-ionized gas is that the Ne$^{+3}$ and Ne$^{+4}$ electron temperatures that we have determined for 0828+193 (a few ×10$^4$ K: Table 7) are far below those predicted by the DS96 pure shock models (T$_{Ne+3}$ ~10$^5$ K; T$_{Ne+4}$ ~3×10$^5$ K). A different set of model parameters or assumptions will be unlikely to resolve this discrepancy, because in the post-shock cooling gas it is the ionization potential of a species that principally determines the temperature at which it can exist. The O$^{++}$ temperatures we have measured are in most cases inconsistent with the shock models of DS96, in the sense that the observed OIII] λλ1661,1666/ [OIII] λ5008 ratios (Table 7) are lower than the values predicted by the pure shock models (>0.08). Only for the two sources with large error bars (0211-122 and 4C-00.54) is the OIII] λλ1661,1666/ [OIII] λ5008 ratio compatible with the predictions of DS96. The O$^+$ temperature measurement for 4C+40.36 (7700±800 K) is in good agreement with the range predicted by the DS96 models (6000-9000 K).

These models also quite clearly fail to explain the positions of the data in the diagnostic diagrams: the observed [OIII]/Hα, [OIII]/[OII] ratios are observed to be far too large (figure 8x), and for many objects the NV/CIV, CIV/CIII] and CII]/CIII] ratios are considerably lower than these models produce (see figure 8b,l).

There is nevertheless a hint that collisional ionization may contribute to the overall ionization of the EELR. Firstly, in two objects, namely 0211-122 and 4C+40.36, the NV/NIV] ratio is significantly higher than any of the photoionization models are able to produce, but is in good agreement with the values predicted by the pure shock models (figure 8g,h). Second, some of the objects are shifted to the left of the photionization loci in the NIV]/HeII vs. NIV]/CIV diagram (figure 8i,j) but are in much better agreement with the pure shock models. It is important to realise that this result is based on the apparent weakness of just one line (NIV]). We suggest OVI λ1035/HeII λ1085 as another line ratio that may help to diagnose whether shocks make a (fractional) contribution to the ionization of the EELR. This is because OVI λ1035/HeII is expected to be much higher in shock-ionized gas than in AGN photoionized gas (DS96). Unfortunately, both OVI λ1035 and HeII λ1085 fall outside the wavelength coverage of our data.

We find no clear correlation between the NV/NIV] ratio and the presence of independent evidence for jet-cloud interactions (e.g. the degree of kinematic disturbance), though the number of sources for which we have measurements of NV/NIV] is probably too small for this result to be considered statistically significant.

Interestingly, the broad-line regions of many (but not all) high-z quasars show a similar discrepancy when their line spectra are compared against photoionization models (e.g. Dietrich et al. 2003). This leads us to suggest that shocks contribute to the ionization of the BLR in many high-z quasars. Indeed, a recent investigation into the EUV lines of a quasar has provided evidence for shock-ionization in the BLR (Binette et al., in preparation).

It is important to note that these results are not in contradiction to the earlier results from this section: while shocks do indeed fail to explain the [OIII]/Hα, [OII]/Hα, NV/CIV, CIV/CIII] and CII]/CIII] ratios, these ratios do not preclude the presence of a significant quantity of shock-ionized gas.

### 4.2.6 Shocks with photoionized precursor

In the majority of the diagnostic diagrams these models are able to explain the positions of most of the data. This dramatic improvement over the pure shock models is the result of including the photoionized precursor component. The precursor dominates many of the optical lines and contributes substantially to the UV line emission and, therefore, these models show marked similarities to pure photoionization predictions. Indeed, the observed values for the temperature diagnostic OIII] λλ1661,1666/ [OIII] λ5008 ratio are generally consistent with the predictions of the fast shock with precursor models. Again, the O$^+$ temperature predicted by the DS96 shock with precursor models is consistent with the observed value. Since DS96 did not tabulate [NeIV] λ1602 or [NeV] λ1575 fluxes, we are unable to compare the observed Ne$^{+3}$ and Ne$^{+4}$ temperatures against the shock with precursor predictions.

As is the case for the pure shock models, these models readily explain NV/NIV] and NIV]/CIV. However, these models encounter some problems in explaining the positions of the data. In particular, many objects are offset significantly to the left of the model predictions in the [OI]/Hα vs. [OII]/[OIII] diagram (figure 8p) and also in the CIII]/HeII vs. CIV/CIII] diagram (figure 8n).

It is not immediately clear as to how these discrepancies could be resolved. For example, increasing the ionization parameter of the photoionized precursor, to account for an additional ionizing flux from the active nucleus, would allow the discrepancy of the [OI]/Hα, [OII]/[OIII] and [OIII]/Hα ratios with the NV/HeII and NV/CIV ratios to be resolved and would also shift the model loci to the left (i.e. closer to the data) in the [OI]/Hα vs. [OII]/[OIII] diagram (figure 8p), but this would also lead to a decrease in [SII]/Hα, thereby exacerbating another discrepancy.

A discrepancy between diagnostic diagrams has also been reported by Solórzano-Iñarrea, Tadhunter & Axon (2001) for intermediate/high redshift radio galaxies. These authors proposed 'matter bounded shocks' to explain the discrepancy. Although matter bounded shocks would help to resolve the discrepancy of [OI]/Hα, [OII]/[OIII] and [OIII]/Hα with NV/HeII and NV/CIV shown by our sample, and [SII]/Hα would become more discrepant.

**Table 10.** The relationship between $A_v$ and evidence for jet-gas interactions. Columns: [1] source name; [2] redshift; [3] $A_v$ in magnitudes; sources are ordered such that $A_v$ increases towards the bottom of the table; the dashed line separates sources with low $A_v$ from those with high $A_v$; in the cases of 0850-206 and 1303+091 we have derived Av using the ratio between the fluxes of HeII $\lambda$3202 and HeII $\lambda$1640 measured by Solórzano-Iñarrea et al. (2004) [4] percentage polarization of the UV continuum emission (Cimatti et al. 1998; Vernet et al. 2001; Solorzano-Iñarrea et al. 2004); [5] global FWHM of the line emission, measured from HeII $\lambda$1640 or [OIII] $\lambda$5007; [6] parameter indicating whether the velocity profile of the emission lines shows a blue wing (Villar-Martín et al. 2002, 2003; Humphrey et al. 2006; Nesvadba et al. 2006); [7] projected size of the radio source, in kpc (Carilli et al. 1997; Pentericci et al. 2000; Solorzano-Iñarrea et al. 2004). Sources with perturbed line kinematics (blue wings and/or high FWHM) have lower $A_v$. Note also the trend for sources with high UV polarization (i.e. >10 per cent) to have higher $A_v$. Both trends can be explained in terms of destruction of dust by jet-induced shocks (see text). *Results for 0850-206 and 1303+091 were taken from Solorzano-Iñarrea et al. (2004); these authors did not measure the global FWHM or present line velocity profiles.

| Source | z | $A_v$ | $P_{UV}$ | Global FWHM | Blue wing? | Radio size |
|--------|-----|---------|---------|--------|--------|--------|
| [1] | [2] | [3] | [4] | [5] | [6] | [7] |
| 0828+193 | 2.57 | ~0 | 10±1 | 1400 | 1 | 98 |
| 1558-003 | 2.57 | ~0 | - | 700 | 1 | 71 |
| 4C+40.36 | 2.27 | ~0 | 7±1 | 1700 | 1 | 32 |
| 2104-242 | 2.49 | <0.4 | - | 700 | 1 | 177 |
| 0406-244 | 2.44 | <0.5 | - | 1800 | - | 82 |
| 1303+091* | 1.41 | <0.6 | 6±1 | - | - | 73 |
| 1138-262 | 2.12 | <0.7 | - | 1000 | 1 | 133 |
| 4C+23.56 | 2.48 | 0.9±0.4 | 15±2 | 700 | 0 | 411 |
| 0850-206* | 1.34 | 1.4±0.1 | 23±3 | - | - | 118 |
| 4C-00.54 | 2.36 | 1.5±0.5 | 12±3 | 800 | 0 | 189 |
| 0529-549 | 2.57 | 1.6±1.0 | - | 800 | 0 | 10 |

Another solution might be to increase the luminosity of the precursor sufficiently for the overall [OI], [OII], [SII], [OIII], Hα and HeII fluxes (i.e. shock plus precursor emission) to be dominated by the precursor photoionized gas.

### 4.2.7 AGN-photoionization with shocks

We have shown that AGN photoionization explains well the data in most of the diagnostic diagrams but cannot explain the ratios involving NIV]. On the other hand, in this section it has been shown that the shock with precursor models are able to explain well the data in some of the diagrams (including those involving NIV]), but appear to require an additional source of ionizing energy. These apparently contradictory results can be reconciled if photoionization by the active nucleus is the dominant ionization mechanism but collisionally-ionized gas is also present in a comparatively small quantity (~10 per cent of the HeII or Hα emission). Fast, cooling shocks can emit NV and CIV very strongly (i.e. NV/HeII~10, CIV/HeII~19 for a 150 km s$^{-1}$ shock; see DS96) and, therefore, even when photoionization is the dominant ionization mechanism and dominates the majority of the emission lines, nearly all of the NV and CIV emission could still in fact originate from shock-ionized gas. The majority of line ratios would then be characteristic of AGN-photoionized gas, while the NIV]/CIV and NV/NIV] ratios would be characteristic of shock-ionized gas, as we find for this sample.

Villar-Martín et al. (2003) have estimated the total mass of gas that would need to be consumed by shocks in order to power the EELR over the expected lifetime of the radio source (~10$^7$ yr) – the required gas masses are an order of magnitude higher than the masses implied by the observed Lyα luminosities. We now consider whether it is energetically feasible that shock ionization is responsible for producing ~10 per cent of the HeII or Hα emission. Assuming a shock speed of 500 km s$^{-1}$, the mass flow rates required to explain 10 per cent of the Hα luminosities of our sample (10$^{42}$ - 10$^{43}$ erg s$^{-1}$) are in the range 10$^2$ - 10$^3$ M$_{sol}$ yr$^{-1}$. A total gas mass of ~10$^9$ – 10$^{10}$ M$_{sol}$ would need to be consumed by shocks, over the lifetime of the radio source (~10$^7$ yr). This is in agreement with the masses of ionized gas implied by the Hα luminosities of our sample (~10$^9$ – 10$^{10}$ M$_{sol}$). Thus, it is energetically feasible that shocks make a fractional contribution to the ionization of the EELR.

## 5. COMPARISON WITH LOW-REDSHIFT RADIO GALAXIES

In order to compare the emission line spectra of HzRG and low-redshift radio galaxies (LzRG hereinafter), several optical diagnostic diagrams have been replotted with the data of Robinson et al. (1987) added. These are shown in figure 10. In all three diagnostic diagrams the HzRG and LzRG occupy similar positions, with no systematic difference being apparent between the two samples. This suggests that the physical conditions, such as ionization state and metallicity, are not dependent on redshift.

One of the LzRG modeled by Robinson et al., namely 2158-380, has also been observed with the International Ultraviolet Explorer by Fosbury et al. (1982). The UV-optical spectrum of this particular object may not necessarily be 'typical' of LzRG but, even so, it affords a useful comparison between the UV-optical emission line spectra of HzRG and LzRG. There appears to be no significant difference between the UV-optical line spectrum of this object and those of the HzRG. Again, this suggests that the physical conditions prevalent in the EELR of powerful LzRG and this sample of HzRG are similar.

## 6. POSSIBLE TREND WITH $A_V$

In previous papers our group has identified a set of trends between a number of the observed properties of this sample (Vernet et al. 2001; Humphrey et al. 2006). To summarize these trends, sources with smaller radio sizes show:
- lower ionization state
- higher UV line luminosities
- higher line FWHM
- lower UV continuum polarization
- a more luminous young stellar population (YSP)

A similar trend between radio size, ionization state and line kinematics has also been reported at z~1 by Best, Röttgering & Longair (2000) and Inskip et al. (2002), and a possible correlation between radio size and polarization was also reported by Solorzano-Iñarrea et al. (2004). We consider the trend between radio size, line FWHM and ionization state to be the result of interactions between the radio source and the ambient ISM: as the radio source propagates outward through the host galaxy, it perturbs and compresses gas clouds in the ISM (see Best, Röttgering & Longair 2000, and Humphrey et al. 2006, for more detailed discussions).

In addition, we find that sources with high $A_v$ do not show evidence for jet-gas interactions. Conversely, sources which do show evidence for jet-gas interactions have low $A_v$. We also note that sources with high UV polarization (i.e. >10 per cent) show high values for $A_v$, while sources with low polarization have low $A_v$. We illustrate this possible trend in Table 10. For our $z\sim2.5$ radio galaxies, $A_v$ was determined from recombination lines of HI and HeII (§3.2). In this table we also include the two radio galaxies studied by Solorzano-Iñarrea et al. (2004), namely 0850-206 ($z$=1.34) and 1303+091 ($z$=1.41); for these two sources we derived $A_v$ using the ratio between HeII λ3203 and HeII λ1640.

De Young (1998) presented calculations showing that shock waves resulting from jet-gas interactions are able to destroy dust in significant quantities. We propose that jet-induced destruction of dust is taking place in some HzRG. This phenomenon would lead to lower $A_v$, and hence higher apparent line luminosities, in sources undergoing strong jet-gas interactions. Similarly, dust destruction could lead to apparently more luminous YSP, which in turn would result in lower polarization due to the dilution of scattered light with more starlight. The destruction of dust might also lead to lower polarization because there would then be less dust to scatter the nuclear light.

An interesting consequence of this scenario is that, if this destruction of dust occurs preferentially along the path of the radio jets, one might expect the line emission and also the YSP to appear closely aligned with the radio axis, as a result of lower dust extinction. For sources showing a close alignment between the radio source and the YSP (e.g. Dey et al. 1997), this effect might provide an alternative to jet-induced star-formation.

## 7. DISCUSSION AND CONCLUSIONS

In this paper we have presented deep, long-slit NIR spectroscopy of the rest-frame optical emission from nine radio galaxies at $z\sim2.5$, obtained at the VLT using the ISAAC instrument. We have also presented deep, long-slit optical spectroscopy for the $z\sim2.5$ radio galaxies 0406-244 and 2025-218, obtained using the LRISp instrument at the Keck II telescope, which samples the rest-frame UV emission.

Wherever possible, we have used broad band photometry to cross-calibrate the flux scales of the optical and NIR spectra in order to produce line spectra spanning a rest wavelength of ~1200-7000 Å. We have also presented a composite line spectrum, running from Lyα through to [SII] λλ6718,6733, for powerful radio galaxies at $z\sim2.5$.

The main results of this paper are as follows.

(i)     We have detected the strong lines [OIII] λ4960, [OIII] λ5008 and Hα λ6565 from all sources in our NIR sample. In almost all cases, the [OII] λλ3728,3729 and [NII] λ6585 lines are detected. From relatively few sources, we also detect the weaker lines [NeV] λ3426, [NeIII] λ3870, Hβ λ4861, [OI] λ6302 and [SII] λλ6718,6733. There is a substantial source-to-source variation in many of the emission line ratios.

(ii)    Very broad Hα (12000 km s$^{-1}$) is detected from 1558-003, which reveals this source to be a broad-line radio galaxy. We have also detected very broad Hα from 1138-262 and 2025-218, in line with previous studies of these two sources

(Nesvadba et al. 2006; Villar-Martín et al. 1999b; Larkin et al. 2000).

(iii)   We have investigated the relative strengths of Lyα, Hβ, Hα, HeII λ1640 and HeII λ4687, and we find that $A_v$ varies significantly from object to object. Several sources are unreddened. We suggest that HeII λ1640/Hβ can be used to make a first order estimate of $A_v$ when Hα/Hβ and HeII λ1640/HeII λ4687 are not available.

(iv)    In addition, we have identified several new line ratios to calculate electron temperature: [NeV] λ1575/[NeV] λ3426, [NeIV] λ1602/[NeIV] λ2423, OIII] λ1663/[OIII] λ5008 and [OII] λ2471/[OII] λ3728. Using OIII] λ1663/[OIII] λ5008 we have calculated an average $T_e$ of 14100±800 K.

(v)     Our detection of the quiescent Lyα halo of 2104-242 (Villar-Martín et al. 2003; Overzier et al. 2001) in the light of [OIII] λ5008 and Hα shows that this halo is ionized, and is not merely neutral gas acting as a Lyα mirror. Similarly, by way of detection of [OII] λ3728, [OIII] λ5008 and Hα, we have shown that the giant Lyα halo of 4C+10.48 (see van Ojik et al. 1997) is ionized.

We have investigated the metallicity and ionization properties of a sample of 16 powerful radio galaxies at $z$>2 using emission line spectra which, spanning Lyα through to [SII] λλ6718,6733, provide a very large range of ionization states and chemical elements. We have concluded that the data are best explained by AGN-photoionization with U varying between objects. This conclusion differs from that of Vernet et al. (2001), who found that U does not vary significantly in this sample. For the UV lines analysed by Vernet et al., there is a degeneracy between U and metallicity. However, having access to the rest-frame optical lines has allowed us to break this degeneracy. In common with several previous investigations we find that single slab photoionization models are unable to reproduce satisfactorily the high ionization and low ionization emission lines simultaneously: higher ionization lines imply higher U than do lower ionization lines, suggesting that the EELR comprise populations of clouds with different effective U. A combination of two or more single slab photoionization models, or alternatively the mixed-media photoionization models of Binette, Wilson & Storchi-Bergmann (1996), present a likely solution to this problem. We note that in either case, a variation in U is required to explain the variation in line ratios from object to object.

While the shock with precursor models can reproduce the data in many diagrams, and moreover explain the NV/NIV] and NIV]/CIV ratios that are problematic for AGN-photoionization models, the shock models are, overall, worse than photoionization models at reproducing the data. For these shock models to provide a satisfactory explanation for the data, an additional source of ionizing photons, presumably the active nucleus, is required. Therefore, we conclude that photoionization is the dominant ionization mechanism for this sample. The main problems encountered by the photoionization and by the shock plus precursor models can likely be resolved if the EELR is mainly photoionized, with an internal range in U, but has a fractional contribution from shock-ionization.

It has not been possible to completely disentangle abundance and ionization effects, but nevertheless it can be said that N/H

is near its solar value if, as we conclude above, photoionization is the dominant ionization mechanism in the EELR. This adds further weight to an earlier conclusion reached by Vernet et al. (2001), who found that roughly solar or super metallicities are required in some of this sample. Our metallicities are much higher than previously reported by Iwamuro et al. (2003), who concluded that the EELR of HzRG have ~0.2 times solar metallicity. This difference is likely to be due to the fact that Iwamuro et al. used a narrower subset of emission lines in their analysis.

We also find that N/H and metallicity do not vary by more than a factor of 2 within our sample. Contrariwise, Vernet et al. (2001) concluded that our sample forms a sequence in metallicity, with metallcity ranging from 0.4 to 4 times solar. As we stated above, the reason for this difference lies in our use of the rest-frame optical emission lines together with the UV lines, which has allowed us to break the degeneracy between metallicity and U that affects the UV lines. (See, for example, the anticorrelation between NV/HeII and [NII]/Hα in table 9). We have also found that there is no significant difference between our sample of HzRG and the low-z radio galaxies of Robinson et al. (1987), in terms of either metallicity or the ionization properties.

We have revisited the trends identified in previous papers by our group (Vernet et al. 2001; Humphrey et al. 2006) between line FWHM, ionization state, line luminosity, radio source size, UV polarization and the luminosity of the young stellar population (YSP). In this paper we find that sources with high $A_v$ do not show evidence for jet-gas interactions, while sources undergoing strong jet-gas interactions have low $A_v$. We also find that sources with high UV polarization (i.e. >10 per cent) show high values for $A_v$. The trend between radio size, line FWHM and ionization state is the result of interactions between the radio source and the ambient ISM.

We conclude that these trends are the result of interactions between the radio source and the ambient ISM. As the radio source propagates outward through the host galaxy, it perturbs, compresses and shocks gas clouds in the ISM (e.g. Best, Röttgering & Longair 2000, and Humphrey et al. 2006), with the result that the EELR of smaller radio sources are observed to have higher line FWHM and lower ionization parameter U. We propose that shocks resulting from jet-gas interactions destroy dust (De Young et al. 1998), which would lead to lower $A_v$, brighter line emission, brighter YSP, and lower UV polarization in radio sources that are undergoing strong jet-gas interactions.

There is a growing body of evidence to suggest that the massive galaxies which become hosts of powerful AGN are assembled very early in the history of the universe. After the original mass build-up such galaxies are thought to evolve fairly passively, as evidenced by the tightness of the correlation in the K – z diagram (e.g. Lilly & Longair 1984; De Breuck et al. 2002), and also by the small scatter in the stellar masses derived from *Spitzer* photometry of radio galaxies across 1 < z < 4 (Seymour et al. 2007). It is thought that during this (mostly) passive evolution, merger events trigger episodes of nuclear activity. This nuclear activity can have a profound effect on the observed properties and, perhaps, on the future evolution, of the host galaxy (Nesvadba et al. 2006). It is likely that these mergers also trigger star formation, the presence of which is implied by the UV continuum properties and/or the luminous sub-millimetre emission of some HzRG (Dey et al. 1997; Reuland et al. 2004). The high rate of star formation at the time of the *original* mass build-up is likely to be responsible for the enrichment to roughly solar metallicity, and also for the dispersal of the enriched gas into the halo by means of starburst powered winds. The high metallicity we have found in the EELR of our HzRG, and also the apparent lack of metallicity variation both from object-to-object and from z~2.5 to z~0, are consistent with this picture.

## ACKNOWLEDGEMENTS

AH acknowledges support from a PPARC studentship, a University of Hertfordshire post-doctoral research assistantship and a UNAM postdoctoral fellowship. MV-M has been supported by the Spanish Ministerio de Educación y Ciencia and the Junta de Andalucía through grants AYA2004-02703 and TIC114. LB acknowledges support from CONACYT through grant J-50296. RF is affiliated to the Research and Science Support Division of the European Space Agency. We would also like to acknowledge the important contribution Marshall Cohen has made to this project. We thank Fumihide Iwamuro and Kentaro Motohara for allowing us to extract 1D spectra from their Subaru 2D frames. We also thank Huub Röttgering for providing information about the spectra in Röttgering et al. (1997). We acknowledge the suggestions made by the anonymous referee that improved this paper.

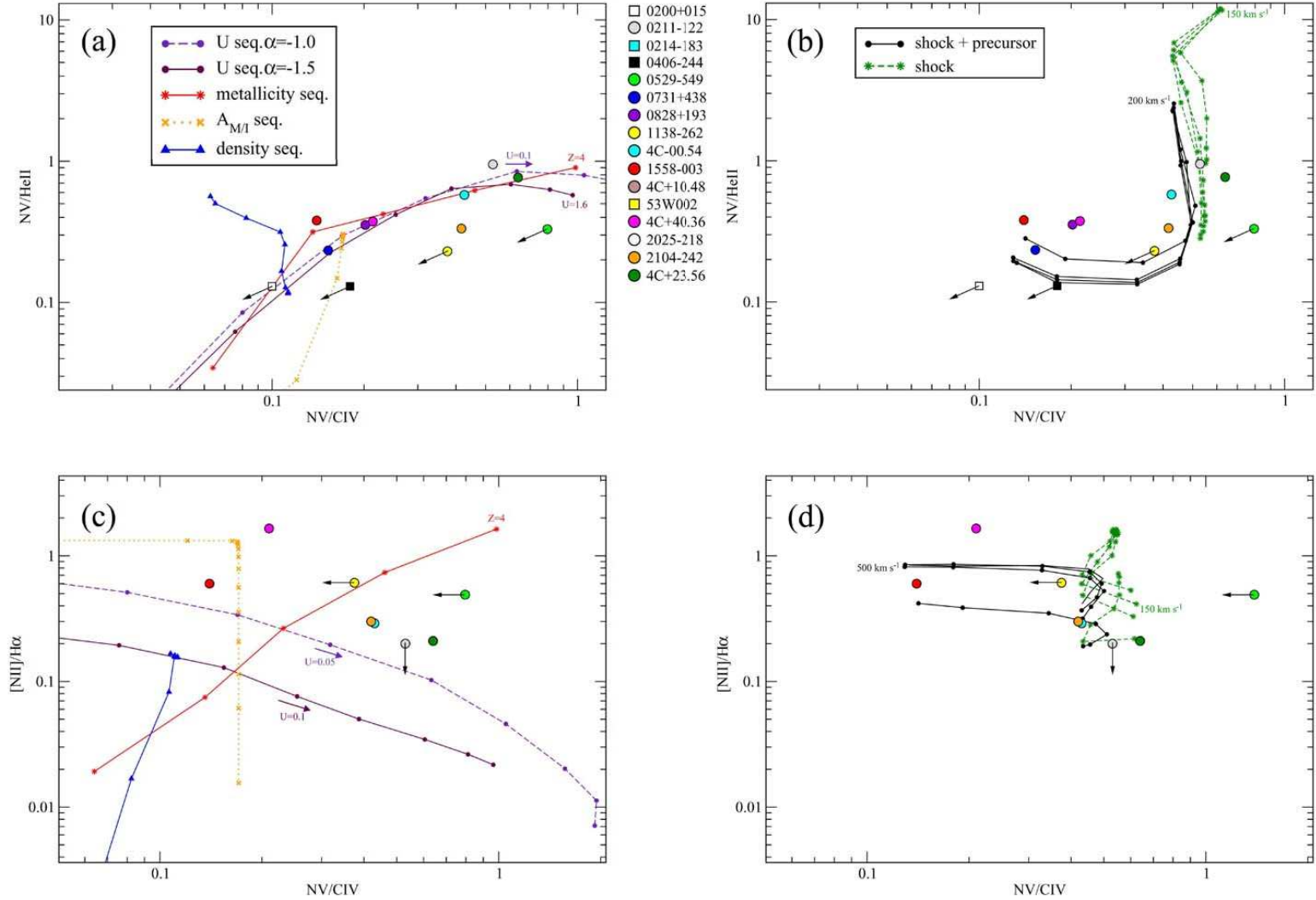

**Figure 8.** Diagnostic diagrams for our sample of HzRG. See text for a description of the ionization models. Left panels: comparison of the data against photoionization models. Right panels: comparison of the data against shock models. Models are represented by lines of different style and colour. Line ratios measured for high-z radio galaxies are represented by symbols of different colour and shape.

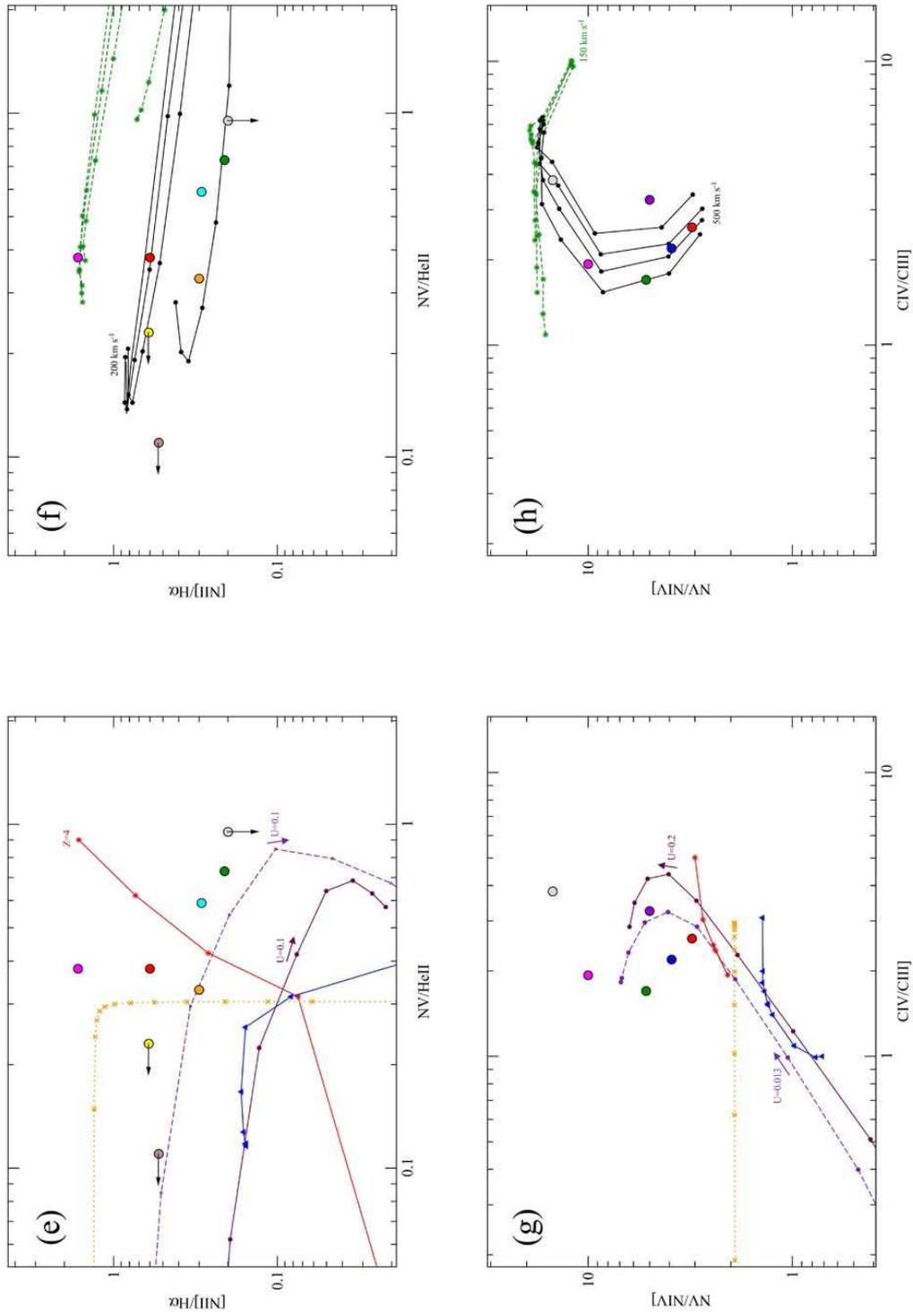

**Figure 8.** Continued.

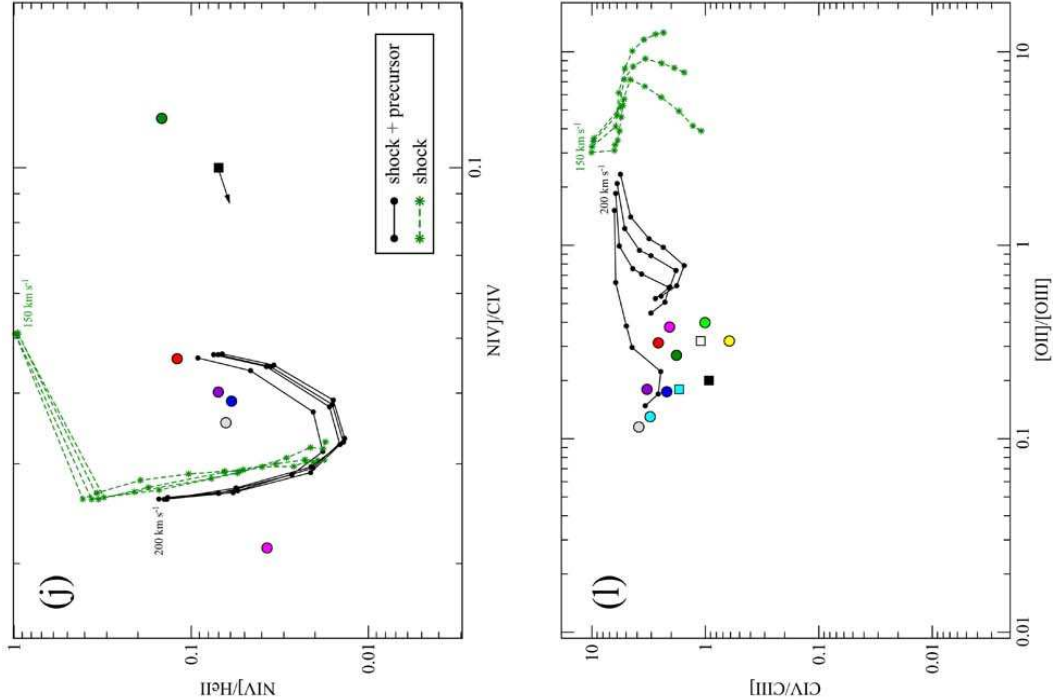

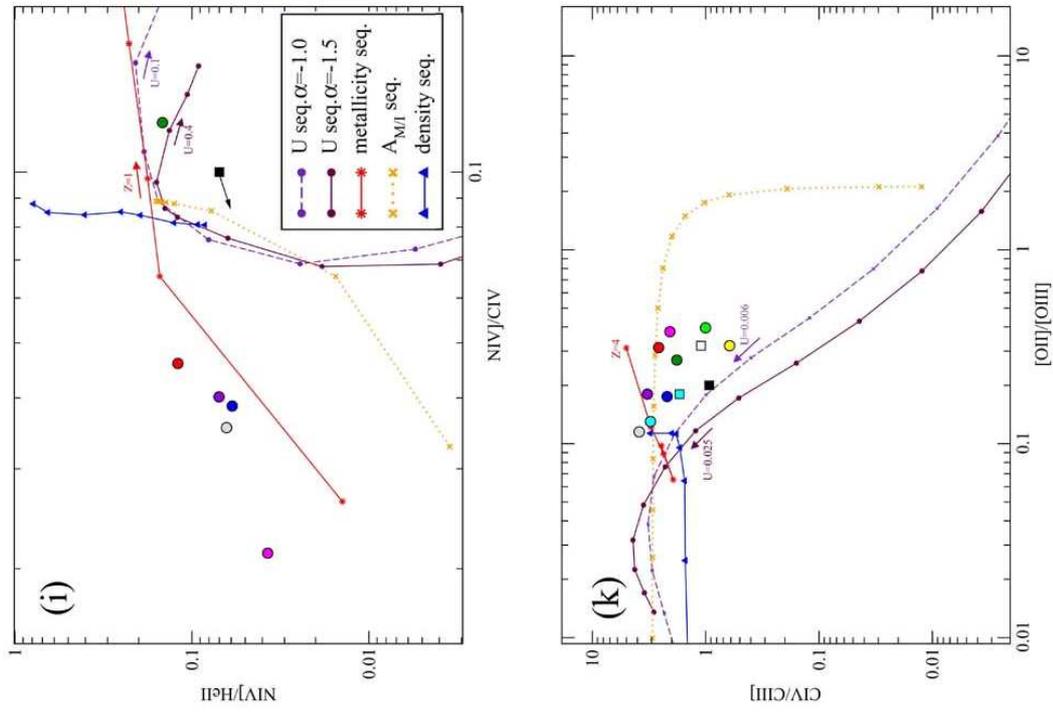

Figure 8. Continued.

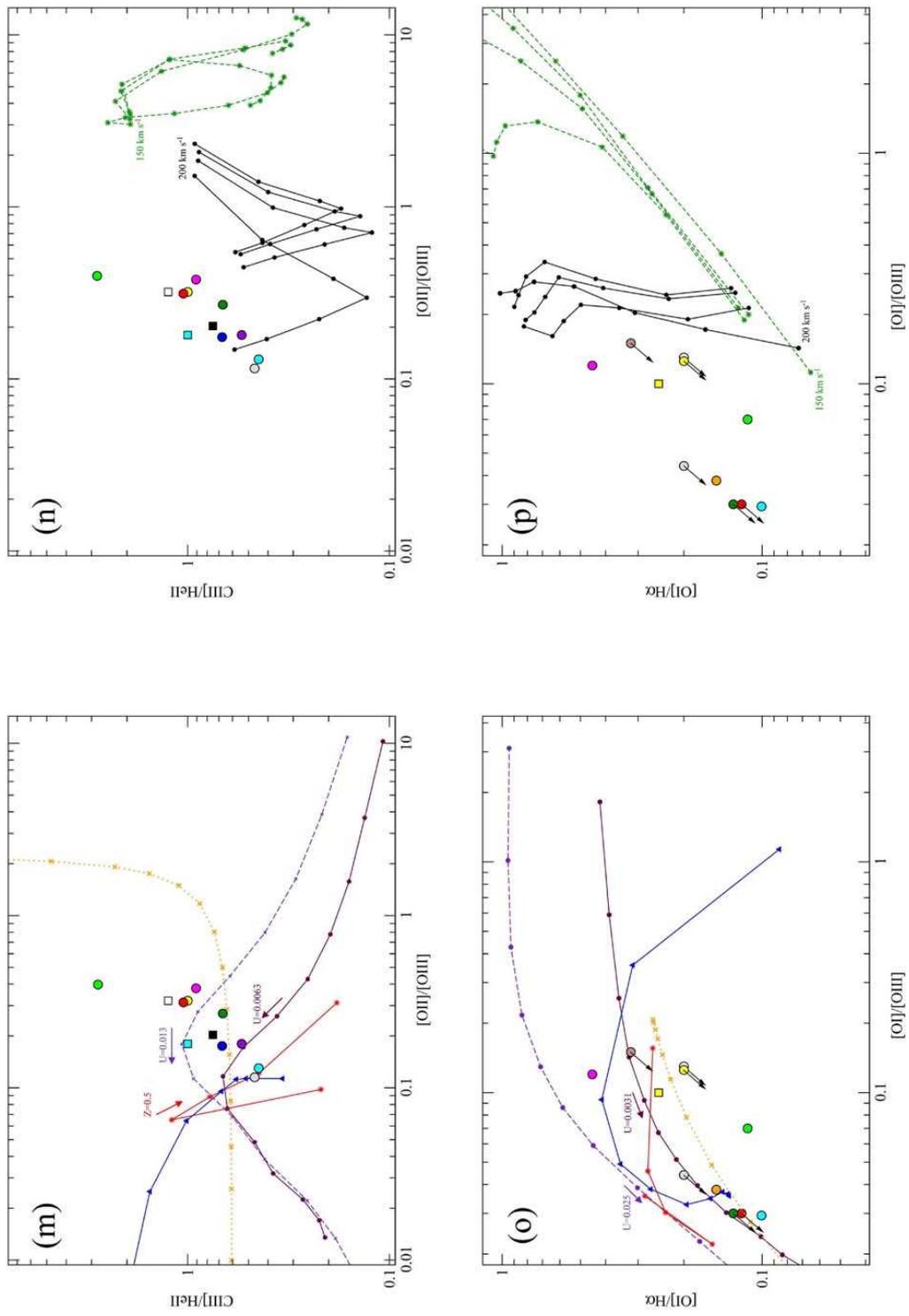

**Figure 8.** Continued.

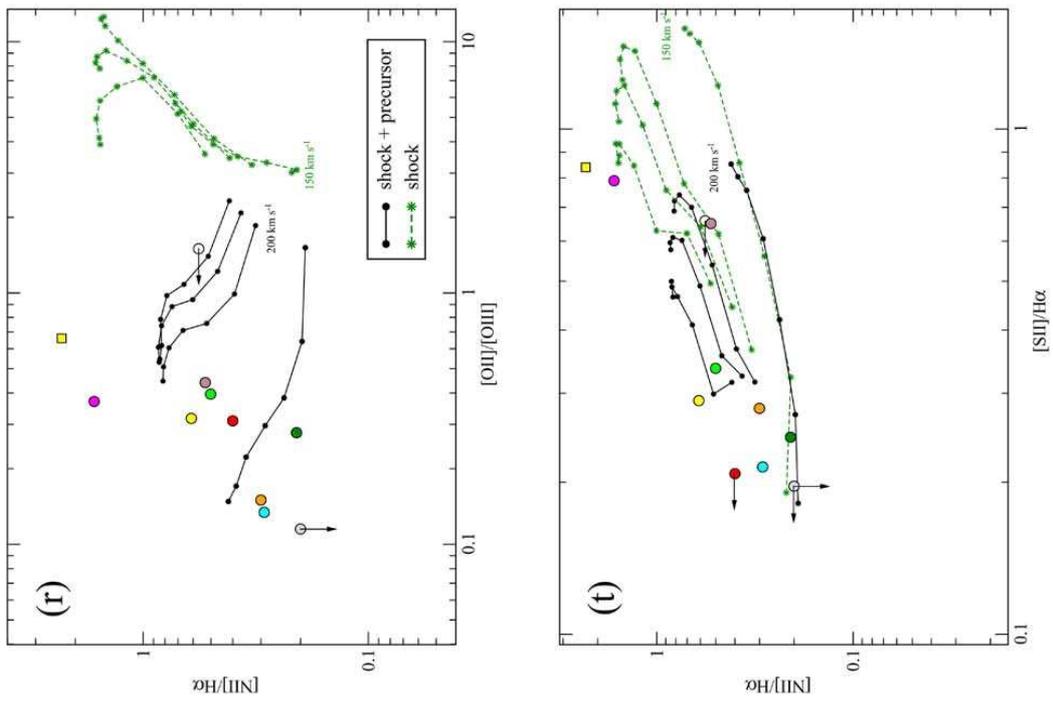

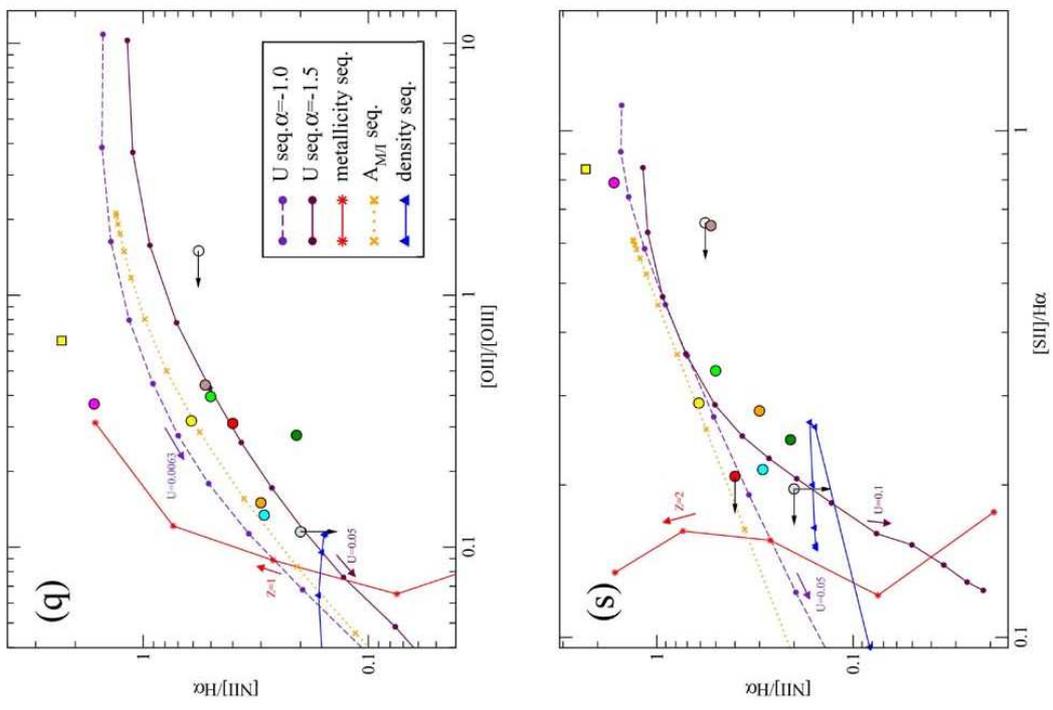

**Figure 8.** Continued.

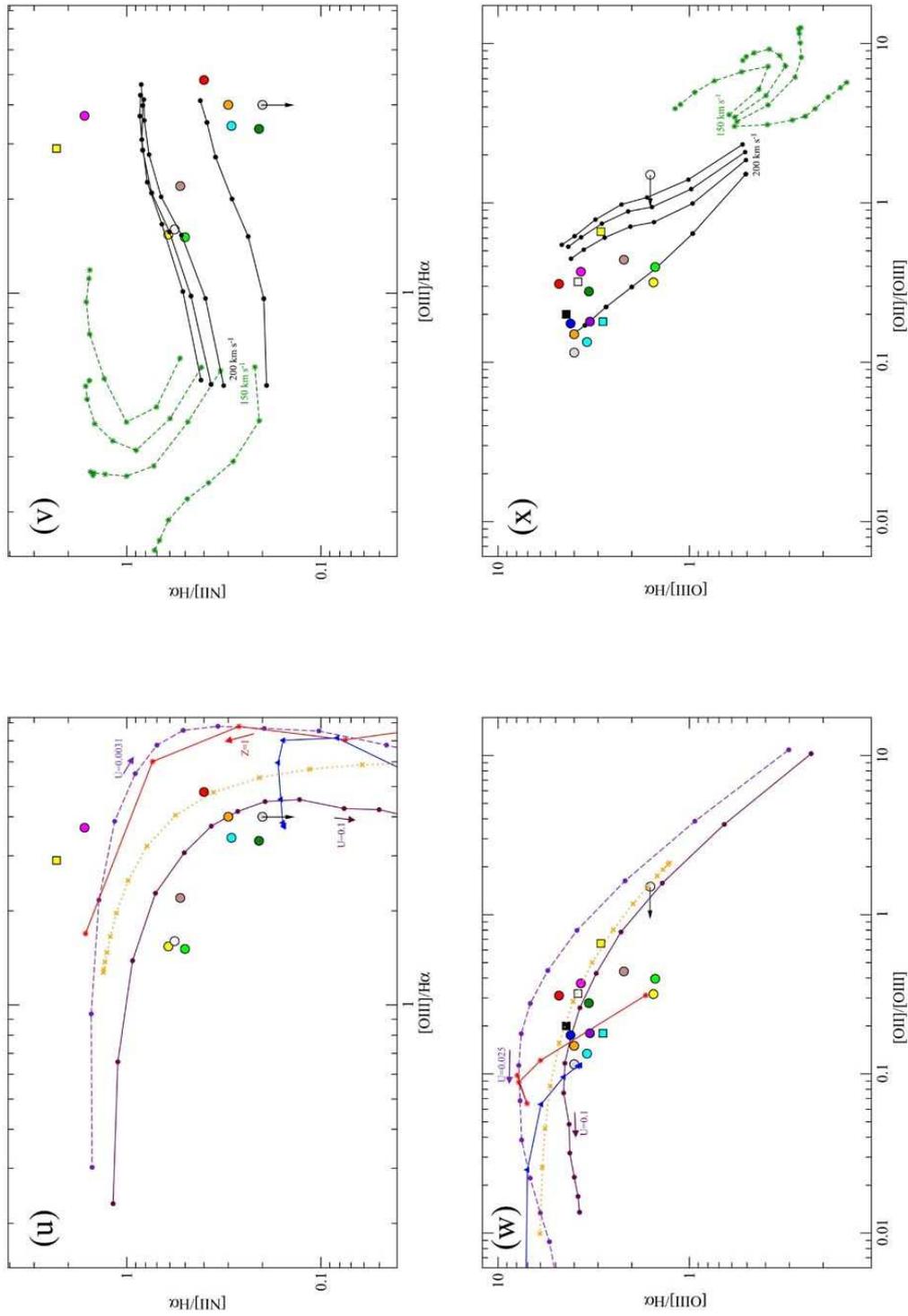

**Figure 8.** Continued.

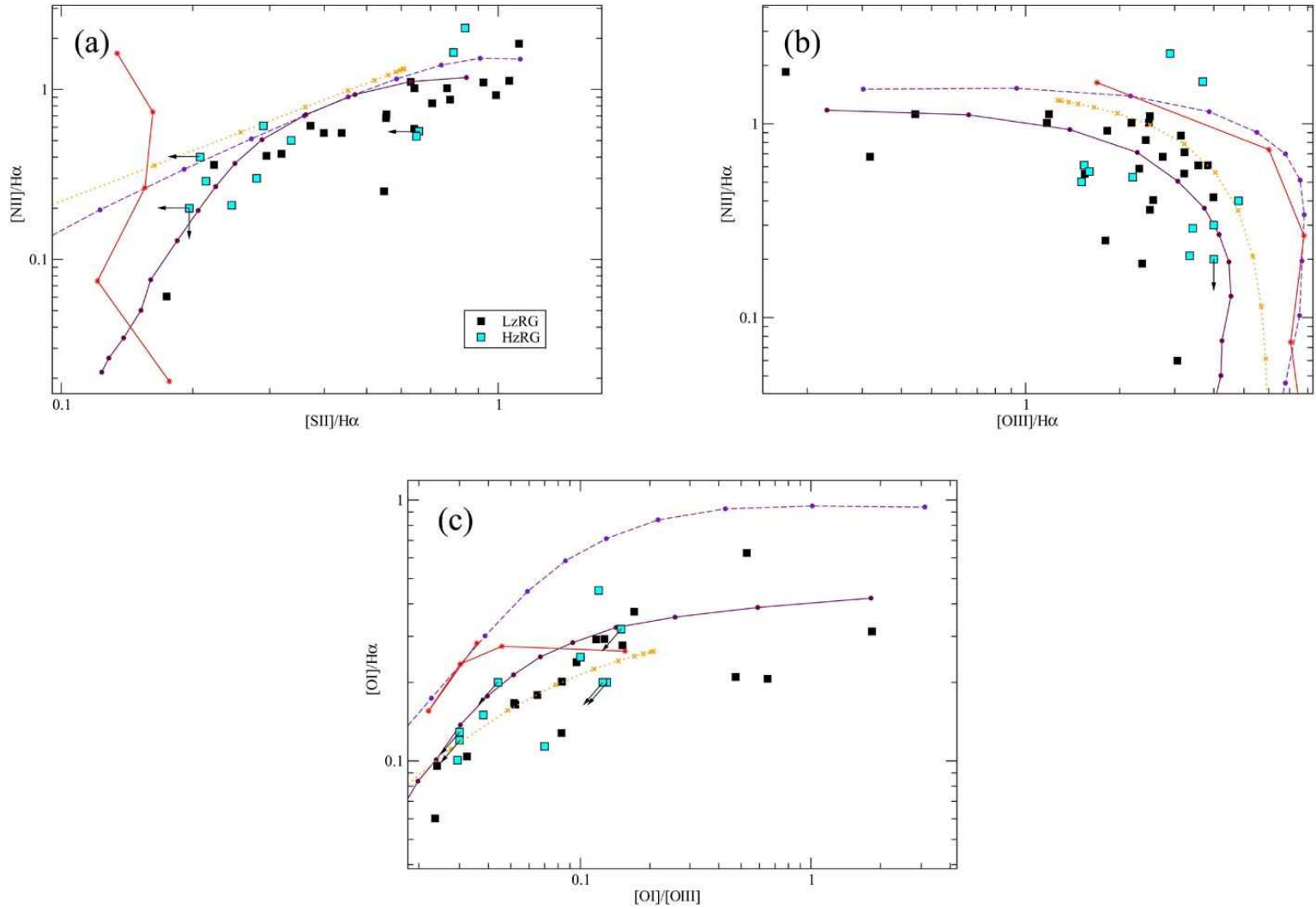

**Figure 9.** Diagnostic diagrams showing low-z radio galaxies and our HzRG sample. The photoionization model loci are the same as those used in Figure 8.

**Table 4.** Measurements of the rest-frame optical emission lines from our sample of HzRG. Where we have been able to perform a cross calibration between the optical and NIR spectra, we also show the rest-frame UV line fluxes. [1] source name   [2] flux of narrow Hα (i.e. excluding any flux from the BLR) in units of 10$^{-16}$ erg s$^{-1}$ cm$^{-2}$ [3] Line [4] vacuum wavelength in Å; for blends of more than two lines, the wavelength is shown in italics [5] line flux, relative to Hα [5] FWHM of line in km s$^{-1}$ [6] velocity shift of line relative to [OIII] λ5008, in km s$^{-1}$ ; for sources with two apertures, velocities are given relative to that of [OIII] in aperture 1. An asterisk (*) is shown next to values for which constraints from another line were used.

| Source [1] | F$_{Hα}$ [2] | Line [3] | λ [4] | Flux [5] | FWHM [6] | Δv [7] |
|---|---|---|---|---|---|---|
| 0211-122 | 6.4±0.5 | Lyα | 1216 | 0.21±0.01 | 1000±100 | |
| | | NV | 1241 | 0.35±0.03 | | |
| | | SiII | 1309 | 0.021±0.002 | | |
| | | OIV], SiIV | *1398* | 0.074±0.008 | | |
| | | NIV] | 1485 | 0.024±0.005 | | |
| | | CIV | 1550 | 0.65±0.02 | | |
| | | HeII | 1640 | 0.36±0.01 | 600±100 | |
| | | OIII] | 1663 | 0.028±0.003 | | |
| | | CIII] | 1908 | 0.17±0.02 | | |
| | | [NeIV] | 2423 | 0.11±0.01 | | |
| | | | | | | |
| | | [NeV] | 3426 | <0.3 | | |
| | | [OII] | 3728 | 0.39±0.08 | 730±160 | 60±150 |
| | | [NeIII] | 3870 | 0.31±0.08 | 390±110 | 220±90 |
| | | [OIII] | 4364 | <0.4 | | |
| | | HeII | 4687 | <0.2 | | |
| | | Hβ quiescent | 4863 | <0.5 | | |
| | | [OIII] quiescent | 4960 | 0.73±0.03* | <500* | 0±8* |
| | | [OIII] quiescent | 5008 | 2.08±0.08 | <500 | 0±8 |
| | | [OIII] perturbed | 4960 | 0.42±0.02* | 1310±60* | -590±40* |
| | | [OIII] perturbed | 5008 | 1.19±0.04 | 1310±60 | -590±40 |
| | | [OI] | 6302 | <0.2 | | |
| | | [NII] | 6550 | <0.04* | | |
| | | Hα quiescent | 6565 | 0.77±0.15 | <500 | 50±20 |
| | | Hα perturbed | 6565 | 0.23±0.08 | 1310±60* | -590±40* |
| | | [NII] | 6585 | <0.2 | | |
| | | [SII] | 6725 | <0.2 | | |
| | | | | | | |
| 0406-244 | Hβ=3.3±0.2 | Lyα | 1216 | 7.0±0.4 | 2100±200 | |
| | | NV | 1241 | <0.2 | | |
| | | OIV], SiIV | *1398* | 0.3±0.1 | | |
| | | NIV] | 1485 | <0.1 | | |
| | | CIV | 1550 | 0.9±0.1 | | |
| | | HeII | 1640 | 1.2±0.1 | 1800±500 | |
| | | OIII] | 1663 | 0.2±0.1 | | |
| | | CIII] | 1908 | 0.9±0.1 | | |
| | | CII] | *2326* | 0.6±0.1 | | |
| | | [NeIV] | 2423 | 1.0±0.3 | | |
| | | | | | | |
| | | [NeV] | 3426 | 0.5±0.1 | | |
| | | [OII] | 3728 | 2.6±0.2 | | |
| | | [NeIII] | 3870 | 1.0±0.1 | | |
| | | HeII | 4687 | 0.21±0.06 | | |
| | | Hβ | 4863 | 1.00±0.07 | | |
| | | [OIII] | 5008 | 12.6±0.1 | | |
| | | | | | | |
| 0529-549 | 10±1 | [NeV] | 3426 | <0.4 | | |
| | | [OII] | 3728 | 0.61±0.19 | 890±180 | 180±50 |
| | | [NeIII] | 3870 | <0.4 | | |
| | | [OIII] | 4364 | <0.2 | | |
| | | HeII | 4687 | <0.2 | | |
| | | Hβ | 4863 | 0.19±0.04 | 680±30* | -580±90* |
| | | [OIII] | 4960 | 0.53±0.01* | 760±20* | 0±10* |
| | | [OIII] | 5008 | 1.51±0.04 | 760±20 | 0±10 |
| | | [OI] | 6302 | 0.11±0.04 | 680±30* | 530±270 |
| | | [NII] | 6550 | 0.17±0.02* | 680±30* | -510±60* |
| | | Hα | 6565 | 1.0±0.1 | 680±30 | -580±90 |
| | | [NII] | 6585 | 0.49±0.08 | 680±30* | -510±60 |
| | | [SII] | 6725 | 0.3±0.1 | 600±40 | -1500±500 |

| Source [1] | F$_{H\alpha}$ [2] | Line [3] | $\lambda$ [4] | Flux [5] | FWHM [6] | $\Delta v$ [7] |
|---|---|---|---|---|---|---|
| 0828+193 | H$\beta$=5.3±0.3 | Ly$\alpha$ | 1216 | 21.7±0.3 | 1200±100 | |
| | | NV | 1241 | 0.9±0.1 | | |
| | | SiII | 1309 | 0.074±0.007 | | |
| | | CII | 1335 | 0.082±0.008 | | |
| | | OIV], SiIV | 1398 | 0.78±0.08 | | |
| | | NIV] | 1485 | 0.18±0.02 | | |
| | | CIV | 1550 | 4.5±0.1 | | |
| | | [NeV] | 1575 | 0.022±0.002 | | |
| | | [NeIV] | 1602 | 0.055±0.005 | | |
| | | HeII | 1640 | 2.57±0.04 | 1400±100 | |
| | | OIII] | 1663 | 0.46±0.05 | | |
| | | NIII] | 1749 | 0.15±0.02 | | |
| | | SiII | 1808 | 0.15±0.02 | | |
| | | SiII | 1817 | 0.035±0.003 | | |
| | | SiIII | 1883 | 0.15±0.02 | | |
| | | SiIII] | 1892 | 0.30±0.03 | | |
| | | CIII] | 1908 | 1.4±0.2 | | |
| | | CII] | 2326 | 0.48±0.05 | | |
| | | [NeIV] | 2423 | 1.2±0.1 | | |
| | | | | | | |
| | | [NeV] | 3426 | 0.76±0.08 | | |
| | | [OII] | 3728 | 1.75±0.06 | | |
| | | H$\gamma$,[OIII] | 4342,4364 | 0.63±0.06 | | |
| | | HeII | 4687 | 0.29±0.05 | | |
| | | H$\beta$ | 4863 | 1.00±0.06 | | |
| | | [OIII] | 5008 | 9.53±0.09 | | |
| 1138-262 | 10.3±1.4 | [OII] | 3728 | 0.49±0.07 | 1000±100 | 260±160 |
| | | [NeIII] | 3870 | 0.28±0.04 | 1110±180 | 1250±140 |
| | | HeII | 4687 | <0.3 | | |
| | | H$\beta$ narrow | 4863 | 0.34±0.07 | 1200±80* | 290±40* |
| | | H$\beta$ BLR | 4863 | <0.9* | | |
| | | [OIII] | 4960 | 0.54±0.04* | 960±50* | 0±20* |
| | | [OIII] | 5008 | 1.5±0.1 | 960±50 | 0±20 |
| | | [OIII] redshifted | 5008 | 0.15±0.05 | 400±60 | 1740±50 |
| | | [OI] | 6302 | <0.2 | | |
| | | [NII] | 6550 | 0.2±0.04* | 1200±80* | 1100±70* |
| | | H$\alpha$ narrow | 6565 | 1.0±0.1 | 1200±80 | 290±40 |
| | | H$\alpha$ BLR | 6565 | 6.8±0.6 | 13900±500 | 2200±170 |
| | | [NII] | 6585 | 0.6±0.1 | 1200±80* | 1100±70 |
| | | [SII] | 6725 | 0.29±0.08 | 1100±100 | 400±500 |
| 4C-00.54 | 10.6±0.8 | Ly$\alpha$ | 1216 | 3.11±0.04 | 1000±100 | |
| | | NV | 1241 | 0.15±0.01 | | |
| | | SiII | 1265 | 0.012±0.004 | | |
| | | CII | 1335 | 0.013±0.004 | | |
| | | OIV], SiIV | 1398 | 0.12±0.01 | | |
| | | CIV | 1550 | 0.35±0.02 | | |
| | | HeII | 1640 | 0.256±0.006 | 800±100 | |
| | | OIII] | 1663 | 0.043±0.004 | | |
| | | CIII] | 1908 | 0.114±0.008 | | |
| | | CII] | 2326 | 0.043±0.004 | | |
| | | [NeIV] | 2423 | 0.27±0.03 | | |
| | | | | | | |
| | | [OII] | 3728 | 0.5±0.1 | | |
| | | [OIII] | 4364 | <0.2 | | |
| | | HeII | 4687 | <0.4 | | |
| | | H$\beta$ | 4863 | <0.2 | | |
| | | [OIII] | 4960 | 1.19±0.02 | 760±30* | 0±10 |
| | | [OIII] | 5008 | 3.43±0.08 | 760±30* | 0±10 |
| | | [OI] | 6302 | 0.10±0.04 | 560±120 | -820±100 |
| | | [NII] | 6550 | 0.10±0.01* | 490±50* | -370±30* |
| | | H$\alpha$ | 6565 | 1.00±0.08 | 510±20 | -140±10 |
| | | [NII] | 6585 | 0.29±0.04 | 490±50 | -370±30 |
| | | [SII] | 6725 | 0.21±0.04 | 460±100 | -80±200 |

**Table 4 continued.**

| Source [1] | F_Hα [2] | Line [3] | λ [4] | Flux [5] | FWHM [6] | Δv [7] |
|---|---|---|---|---|---|---|
| 1558-003 | 2.1±0.6 | Lyα | 1216 | 13.9±0.3 | 1200±100 | |
| | | NV | 1241 | 0.41±0.09 | | |
| | | OIV], SiIV | *1398* | 0.37±0.09 | | |
| | | NIV] | 1485 | 0.13±0.04 | | |
| | | CIV | 1550 | 2.4±0.1 | | |
| | | HeII | 1640 | 0.90±0.03 | 740±140 | |
| | | OIII] | 1663 | 0.39±0.07 | | |
| | | SiIII] | 1888 | 0.12±0.03 | | |
| | | CIII] | 1908 | 0.92±0.03 | | |
| | | CII] | *2326* | 0.4±0.1 | | |
| | | [NeIV] | 2423 | 0.5±0.1 | | |
| | | [NeV] | 3426 | <0.9 | | |
| | | [OII] | 3728 | 1.5±0.2 | 680±90 | 460±40 |
| | | [NeIII] | 3870 | <1.3 | | |
| | | HeII | 4687 | <0.6 | | |
| | | Hβ narrow | 4863 | 0.6±0.2 | 580±30* | 210±70 |
| | | [OIII] | 4960 | 1.7±0.1* | 580±30* | 0±10* |
| | | [OIII] | 5008 | 4.8±0.3 | 580±30 | 0±10 |
| | | [OI] | 6302 | <0.06 | | |
| | | [NII] | 6550 | 0.15±0.07* | 580±30* | -90±190* |
| | | Hα narrow | 6565 | 1.0±0.3 | 580±30* | 410±70 |
| | | Hα BLR | 6565 | 7.5±2.4 | 12000±2000 | -3900±400 |
| | | [NII] | 6585 | 0.7±0.2 | 580±30* | -90±190 |
| | | [SII] | 6725 | <0.2 | | |
| 4C+10.48 Ap1 + Ap2 | 6.7±0.8 | [NeV] | 3426 | <1.0 | | |
| | | [OII] | 3728 | 1.0±0.2 | | |
| | | [NeIII] | 3870 | <0.7 | | |
| | | [OIII] | 4364 | <0.4 | | |
| | | Hβ | 4863 | <0.4 | | |
| | | [OIII] | 4960 | 0.75±0.06* | | |
| | | [OIII] | 5008 | 2.2±0.2 | | |
| | | [OI] | 6302 | 0.4±0.1 | | |
| | | [NII] | 6550 | 0.18±0.04* | | |
| | | Hα | 6565 | 1.0±0.1 | | |
| | | [NII] | 6585 | 0.5±0.1 | | |
| | | [SII] | 6725 | 0.6±0.1 | | |
| 4C+10.48 Ap1 | 3.7±0.7 | [NeV] | 3426 | <1.3 | | |
| | | [OII] | 3728 | 1.1±0.3 | 1540±640 | 630±210 |
| | | [NeIII] | 3870 | <0.7 | | |
| | | [OIII] | 4364 | <0.8 | | |
| | | Hβ | 4863 | <0.6 | | |
| | | [OIII] | 4960 | 0.7±0.1 | 950±70* | 0±30 |
| | | [OIII] | 5008 | 2.0±0.4 | 950±70 | 0±30 |
| | | [OI] | 6302 | <0.5 | | |
| | | [NII] | 6550 | 0.21±0.06* | 760±180* | 380±80* |
| | | Hα | 6565 | 1.0±0.2 | 830±190 | 460±60 |
| | | [NII] | 6585 | 0.6±0.2 | 760±180 | 380±80 |
| | | [SII] | 6725 | 0.9±0.2 | 760±180* | 380±80* |

**Table 4 continued.**

| Source [1] | F_Hα [2] | Line [3] | λ [4] | Flux [5] | FWHM [6] | Δv [7] |
|---|---|---|---|---|---|---|
| 4C+10.48 | 3.0±0.3 | [NeV] | 3426 | <0.8 | | |
| Ap2 | | [OII] | 3728 | 0.8±0.2 | 410±220 | 160±80 |
| | | [NeIII] | 3870 | <0.8 | | |
| | | [OIII] | 4364 | <0.7 | | |
| | | Hβ | 4863 | <0.3 | | |
| | | [OIII] | 4960 | 0.84±0.07 | 500±30* | 0±10* |
| | | [OIII] | 5008 | 2.4±0.2 | 500±30 | 0±10 |
| | | [OI] | 6302 | <0.5 | | |
| | | [NII] | 6550 | 0.14±0.03 | 450±20* | 70±50* |
| | | Hα | 6565 | 1.0±0.1 | 450±20 | -10±30 |
| | | [NII] | 6585 | 0.41±0.08 | 450±20* | 70±50 |
| | | [SII] | 6725 | 0.4±0.1 | 450±20* | 70±50* |
| 4C+40.36 | 19.8 | Lyα | 1216 | 5.8±0.2 | 2400±100 | |
| | | NV | 1241 | 0.15±0.03 | | |
| | | SiII | 1309 | 0.048±0.05 | | |
| | | CII | *1335* | 0.11±0.01 | | |
| | | OIV],SIV | *1398* | 0.11±0.01 | | |
| | | NIV] | 1485 | 0.015±0.003 | | |
| | | CIV | 1550 | 0.69±0.03 | | |
| | | HeII | 1640 | 0.39±0.03 | 1700±100 | |
| | | OIII] | 1663 | 0.063±0.006 | | |
| | | NIII] | *1749* | 0.042±0.004 | | |
| | | SiII | 1817 | 0.051±0.005 | | |
| | | CIII] | 1908 | 0.37±0.06 | | |
| | | CII] | *2326* | 0.22±0.02 | | |
| | | [NeIV] | 2423 | 0.19±0.02 | | |
| | | [OII] | 2471 | 0.11±0.01 | | |
| | | [OII] | 3728 | 1.45±0.08 | | |
| | | [NeIII] | 3870 | 0.32±0.08 | | |
| | | HeII | 4687 | 0.08±0.06 | | |
| | | Hβ | 4863 | 0.34±0.05 | | |
| | | [OIII] | 5008 | 3.69±0.06 | | |
| | | [NII] | 6550 | 0.5 | | |
| | | Hα | 6565 | 1.0 | | |
| | | [NII] | 6585 | 1.6 | | |
| | | [SII] | 6725 | 0.8 | | |
| 2025-218 | 6±2 | Lyα | 1216 | 0.8±0.1 | 800±100 | |
| | | NV | 1241 | 0.10±0.02 | | |
| | | CIV | 1550 | 0.07±0.02 | | |
| | | HeII | 1640 | 0.05±0.01 | 500±100 | |
| | | SiIII | 1888 | 0.09±0.06 | | |
| | | CIII] | 1908 | 0.18±0.06 | | |
| | | [NeV] | 3426 | <0.3 | | |
| | | [OII] | 3728 | <2.4 | | |
| | | [NeIII] | 3870 | <2.4 | | |
| | | HeII | 4687 | <0.3 | | |
| | | Hβ narrow | 4863 | <0.3 | | |
| | | [OIII] | 4960 | 0.6±0.1 | 550±150* | 0±170* |
| | | [OIII] | 5008 | 1.6±0.3 | 550±150 | 0±170 |
| | | [OI] | 6302 | <0.2 | | |
| | | [NII] | 6550 | 0.2±0.1 | 440±150* | 80±300* |
| | | Hα narrow | 6565 | 1.0±0.3 | 440±150 | 80±300 |
| | | Hα BLR | 6565 | 2.4±0.7 | 5600±850 | 2810±470 |
| | | [NII] | 6585 | 0.6±0.2 | 440±150* | 80±300* |
| | | [SII] | 6725 | <0.7 | | |

**Table 4 continued.**

| Source [1] | $F_{H\alpha}$ [2] | Line [3] | $\lambda$ [4] | Flux [5] | FWHM [6] | $\Delta v$ [7] |
|---|---|---|---|---|---|---|
| 2104-242 | 7.1±0.4 | Lyα | 1216 | 13±1 | 610±140 | |
| Ap1 + Ap2 | | NV | 1241 | 0.32±0.03 | | |
| | | OIV], SiIV | *1398* | 0.39±0.03 | | |
| | | CIV | 1550 | 0.8±0.1 | | |
| | | HeII | 1640 | 1.0±0.1 | 700±100 | |
| | | OIII] | 1663 | 0.10±0.03 | | |
| | | [NeV] | 3426 | 0.2±0.1 | | |
| | | [OII] | 3728 | 0.6±0.1 | | |
| | | [NeIII] | 3870 | 0.3±0.1 | | |
| | | HeII | 4687 | <0.2 | | |
| | | Hβ | 4863 | 0.3±0.1 | | |
| | | [OIII] | 4960 | 1.4±0.2 | | |
| | | [OIII] | 5008 | 4.0±0.5 | | |
| | | [OI] | 6302 | 0.12±0.06 | | |
| | | [NII] | 6550 | 0.10±0.02 | | |
| | | Hα | 6565 | 1.0±0.1 | | |
| | | [NII] | 6585 | 0.30±0.05 | | |
| | | [SII] | 6725 | 0.32±0.09 | | |
| 2104-242 | 4.8±0.3 | [NeV] | 3426 | <0.2 | | |
| Ap1 | | [OII] | 3728 | 0.7±0.1 | 1000±90 | -690±100 |
| | | [NeIII] | 3870 | 0.3±0.1 | 1020±150 | -540±70 |
| | | HeII | 4687 | <0.1 | | |
| | | Hβ | 4863 | 0.23±0.06 | 530±20* | -370±20* |
| | | [OIII] | 4960 | 1.13±0.02* | 590±120* | 0±5* |
| | | [OIII] | 5008 | 3.24±0.09 | 590±120 | 0±5 |
| | | [OI] | 6302 | 0.14±0.06 | 530±20* | -370±20* |
| | | [NII] | 6550 | 0.11±0.02* | 530±20* | -830±50 |
| | | Hα | 6565 | 1.00±0.06 | 530±20 | -370±20 |
| | | [NII] | 6585 | 0.34±0.05 | 530±20* | -830±50 |
| | | [SII] | 6725 | 0.28±0.09 | 530±20* | 40±140 |
| 2104-242 | 2.3±0.2 | [NeV] | 3426 | <0.2 | | |
| Ap2 | | [OII] | 3728 | 0.4±0.1 | 650±180 | 20±100 |
| | | [NeIII] | 3870 | <0.2 | | |
| | | HeII | 4687 | <0.2 | | |
| | | Hβ | 4863 | <0.2 | | |
| | | [OIII] | 4960 | 1.52±0.07 | 560±20* | 0±10* |
| | | [OIII] | 5008 | 4.4±0.2 | 560±20 | 0±10 |
| | | [OIII] blueshifted | 4960 | 0.12±0.04* | 60±150* | -3320±30* |
| | | [OIII] blueshifted | 5008 | 0.33±0.09 | 60±150 | -3320±30 |
| | | [OI] | 6302 | <0.08 | | |
| | | [NII] | 6550 | 0.08±0.02* | 280±50* | -680±90* |
| | | Hα | 6565 | 1.0±0.1 | 280±50 | -120±20 |
| | | [NII] | 6585 | 0.25±0.06 | 280±50* | -680±90 |
| | | [SII] | 6725 | <0.4 | | |
| 4C+23.56 | 10.2±0.7 | Lyα | 1216 | 0.76±0.01 | 1100±100 | |
| | | NV | 1241 | 0.11±0.01 | | |
| | | OIV], SiIV | *1398* | 0.018±0.002 | | |
| | | NIV] | 1485 | 0.021±0.006 | | |
| | | CIV | 1550 | 0.18±0.02 | | |
| | | HeII | 1640 | 0.15±0.01 | 700±100 | |
| | | OIII] | 1663 | 0.040±0.004 | | |
| | | CIII] | 1908 | 0.10±0.01 | | |
| | | CII] | *2326* | 0.078±0.008 | | |
| | | [NeIV] | 2423 | 0.13±0.01 | | |
| | | [NeV] | 3426 | 0.26±0.07 | 1070±310 | -680±110 |
| | | [OII] | 3728 | 0.9±0.2 | 1660±280 | 160±90 |
| | | [NeIII] | 3870 | 0.26±0.07 | 950±380 | -1480±110 |
| | | HeII | 4687 | <0.3 | | |
| | | Hβ | 4863 | 0.35±0.15 | 730±230 | 210±110 |
| | | [OIII] | 4960 | 1.17±0.05 | 800±20* | 0±10* |
| | | [OIII] | 5008 | 3.4±0.2 | 800±20 | 0±10 |
| | | [OI] | 6302 | <0.1 | | |
| | | [NII] | 6550 | 0.07±0.02 | 880±70* | 5±30* |
| | | Hα | 6565 | 1.00±0.07 | 880±70 | 5±30 |
| | | [NII] | 6585 | 0.21±0.07 | 880±70* | 5±30* |
| | | [SII] | 6725 | 0.3±0.1 | 880±70* | 5±30* |

**Table 5.** Composite UV-optical line spectrum of HzRG. [1] Line ID. [2] restframe vacuum wavelength of the line. [3] line flux relative to Hβ. [4] number of measurements used to calculate the average flux. Some lines were detected in only one source, but these are included for completeness. [5] line fluxes from the average of McCarthy (1993).

| Line<br>[1] | λ<br>[2] | Flux<br>[3] | N☉<br>[4] | McCarthy 93<br>[5] |
|---|---|---|---|---|
| Lyα | 1216 | 14 | 9 | 31 |
| NV | 1241 | 0.62 | 8 | 1.5 |
| SiII | 1309 | 0.08 | 3 | |
| CII | *1335* | 0.13 | 3 | 0.4 |
| OIV], SiIV | *1398* | 0.48 | 8 | 1.6 |
| NIV] | 1485 | 0.11 | 5 | |
| CIV | 1550 | 2.1 | 9 | 3.6 |
| [NeV] | 1575 | 0.02 | 1 | |
| [NeIV] | 1602 | 0.05 | 1 | |
| HeII | 1640 | 1.2 | 9 | 3.2 |
| OIII] | 1663 | 0.29 | 8 | 0.7 |
| NIII] | *1749* | 0.12 | 2 | |
| SiII | 1808 | 0.14 | 1 | |
| SiII | 1817 | 0.08 | 2 | |
| SiIII] | 1883,1892 | 0.32 | 3 | |
| CIII] | 1908 | 0.87 | 8 | 1.8 |
| CII] | *2326* | 0.49 | 6 | 0.9 |
| [NeIV] | 2423 | 0.74 | 7 | 0.9 |
| [OII] | 2471 | 0.23 | 1 | |
| [NeV] | 3426 | 0.59 | 4 | 0.4 |
| [OII] | 3728 | 2.2 | 8 | 3.6 |
| [NeIII] | 3870 | 0.81 | 5 | 0.8 |
| Hγ,[OIII] | 4342,4364 | 0.58 | 1 | 0.3 |
| HeII | 4687 | 0.22 | 3 | 0.2 |
| Hβ | 4863 | 1.0 | 6 | 1.0 |
| [OIII] | 5008 | 9.3 | 9 | 8.7 |
| [OI] | 6302 | 0.29 | 2 | |
| Hα | 6565 | 2.6 | 7 | |
| [NII] | 6585 | 1.6 | 6 | |
| [SII] | 6725 | 1.1 | 4 | |
| T_{O++} | 14100    ±800 | | 8 | |

33